\documentclass[manuscript]{aastex63}
\usepackage{ragged2e}

\providecommand{\e}[1]{\ensuremath{\times 10^{#1}}}

\submitjournal{apj}

\shortauthors{Salman et al.}

\begin{document}

\title{Properties of the Sheath Regions of Coronal Mass Ejections with or without Shocks from STEREO \textit{in situ} Observations near 1 AU}

\correspondingauthor{Tarik M. Salman}
\email{ts1090@wildcats.unh.edu}

\author[0000-0001-6813-5671]{Tarik M. Salman}
\affiliation{Space Science Center and Department of Physics \\
University of New Hampshire \\
Durham, NH 03824, USA}

\author[0000-0002-1890-6156]{No{\'e} Lugaz}
\affiliation{Space Science Center and Department of Physics \\
University of New Hampshire \\
Durham, NH 03824, USA}

\author[0000-0001-8780-0673]{Charles J. Farrugia}
\affiliation{Space Science Center and Department of Physics \\
University of New Hampshire \\
Durham, NH 03824, USA}

\author[0000-0002-9276-9487]{Reka M. Winslow}
\affiliation{Space Science Center and Department of Physics \\
University of New Hampshire \\
Durham, NH 03824, USA}

\author[0000-0002-6849-5527]{Lan K. Jian}
\affiliation{Heliophysics Science Division \\
NASA Goddard Space Flight Center \\
Greenbelt, MD 20771, USA}

\author[0000-0003-3752-5700]{Antoinette B. Galvin}
\affiliation{Space Science Center and Department of Physics \\
University of New Hampshire \\
Durham, NH 03824, USA}

\begin{abstract}

\justify

We examine 188 coronal mass ejections (CMEs) measured by the twin STEREO spacecraft during 2007-2016 to investigate the generic features of the CME sheath and the magnetic ejecta (ME) and dependencies of average physical parameters of the sheath on the ME. We classify the MEs into three categories, focusing on whether a ME drives both a shock and sheath, or only a sheath, or neither, near 1 AU. We also reevaluate our initial classification through an automated algorithm and visual inspection. We observe that even for leading edge speeds greater than 500 km\,s$^{-1}$, 1 out of 4 MEs do not drive shocks near 1 AU. MEs driving both shocks and sheaths are the fastest and propagate in high magnetosonic solar wind, whereas MEs driving only sheaths are the slowest and propagate in low magnetosonic solar wind. Our statistical and superposed epoch analyses indicate that all physical parameters are more enhanced in the sheath regions following shocks than in sheaths without shocks. However, differences within sheaths become statistically less significant for similar driving MEs. We also find that the radial thickness of ME-driven sheaths apparently has no clear linear correlation with the speed profile and associated Mach numbers of the driver. 

\end{abstract}

\keywords{coronal mass ejections --- sheath regions --- shocks}

\section{Introduction}

\justify

In essence, \textbf{coronal mass ejections (CMEs)} are large-scale transient eruptions of solar plasma and magnetic field into the heliosphere \citep[e.g.,][]{Webb:2012}. CMEs are one of the most spectacular and energetic forms of solar phenomena, with kinetic energies potentially exceeding 10$^{32}$ ergs \citep{Hudson:2006}. The interplanetary manifestation of a fast CME typically involves three distinct features: a shock wave compressing and deflecting the upstream solar wind flow, the region of compressed solar wind bounded by the shock front and the leading edge of the following \textbf{magnetic ejecta (ME)}, termed as the sheath, and a cold, magnetically-dominated region behind the sheath, termed as the ME. However, sometimes the ME is preceded by a dense sheath even in the absence of shocks, for example, the well-studied 12 December 2008 CME \citep[see][]{DeForest:2013} or the 17 January 2013 CME, studied by \citet{Lugaz:2016a}. In addition, databases of CMEs measured near 1 AU \citep{Jian:2018} and at 1 AU \citep{Richardson:2010,Nieves:2018} include CMEs without shocks but with a start time several hours before the beginning of the ME, corresponding to a period of dense and disturbed solar wind which can be identified as the sheath region. Though MEs and their subset magnetic clouds \citep[MCs;][]{Burlaga:1981} are the primary drivers of intense geomagnetic disturbances \citep{Farrugia:1997,Zhang:2007,Richardson:2012}, a substantial amount of moderate-intense geomagnetic disturbances are associated with sheath plasma and magnetic field behind shocks \citep{Kamide:1998b,Richardson:2001,Huttunen:2004,Yermolaev:2010,Echer:2013}. Geoeffective shocks/sheaths \citep[see][]{Ontiveros:2010,Lugaz:2016} provide even shorter lead times for forecasts than MEs. ME-driven shocks and sheaths are significant in solar-terrestrial studies as shocked plasma in the immediate downstream of the shock may contribute in the acceleration of solar energetic particles \citep{Reames:1999,Manchester:2005} and significantly compress and distort the magnetosphere \citep{Joselyn:1990}, enhancing the geoeffective potential of MEs.

\justify

Throughout the rest of this manuscript, we define the sheath associated with a ME as the region preceding the ME and driven by it, which contains compressed plasma and magnetic field, \textbf{regardless of the presence of a shock}. This sheath region is associated with the propagation and expansion of the ME and the accumulation and compression of solar wind plasma and magnetic field in front of it. When the leading edge speed of a ME relative to the ambient solar wind is faster than the fast-magnetosonic speed of the medium, it produces a shock (fast-forward) in the frame of the ME \citep{Russell:2002}. In many cases, MEs also expand at significant speeds, although the expansion is typically slower than the local fast-magnetosonic speed \citep{Klein:1982}. This expansion results in the continuous accumulation of solar wind plasma and magnetic field upstream of the ME \citep{Takahashi:2017}. When there is a fast-forward shock, the region consisting of this compressed and heated plasma with stretched and turbulent magnetic field upstream of the ME is identified as the sheath \citep{Kaymaz:2006}. In this study, we extend this definition of sheaths to MEs that do not drive fast-forward shocks but for which there is compressed plasma and magnetic field upstream of it. While both the sheath and ME feature elevated magnetic field signatures, the sheath typically consists of a rapidly fluctuating magnetic field direction \citep{Hietala:2014}. ME-driven sheaths are often compared to planetary magnetosheaths as the solar wind tries to flow around the magnetic obstacle, just like the solar wind flows around the Earth's magnetosphere. However, as MEs not only propagate through interplanetary medium but also expand into it, sheaths upstream of MEs significantly differ from planetary magnetosheaths \citep{Siscoe:2008,Demoulin:2009}. The reduced lateral deflection of the solar wind away from the nose of the ME may not be enough for it to flow around, resulting in the pile up of solar wind in front of it \citep{Siscoe:2008,Owens:2017a}. These layers are not simple stacks of compressed solar wind along the radial line from the Sun because the layers can also slide laterally by different amounts as they are compressed. As a result, ME-driven sheaths are often associated with complex formations and the sheath field appears to be highly variable and turbulent \citep{Kataoka:2005}, for example, the north/south direction of the magnetic field does not present any clear relationship between the sheath region and ME \citep{Jian:2018}. This can make understanding their global configuration and predicting their geoeffectivity particularly challenging \citep{Palmerio:2016,Kilpua:2017a}.

\justify

Embedded magnetic fields in shock-driven sheath regions have significant potential to impact the geomagnetic field \citep{Tsurutani:1988,Lugaz:2016}. Shock-driven sheath regions are often associated with prolonged periods (several hours) of sustained and strong southward fields. Such out-of-ecliptic fields in the sheath regions are a byproduct of two possible mechanisms: compression of preexisting southward B$_{z}$ at the boundary of a shock with a normal close to the ecliptic \citep{Lugaz:2016} or magnetic field line draping upstream of the ME \citep{McComas:1989,Gonzalez:1994}. Planar magnetic structures, where the magnetic field is ordered into laminar sheets \citep{Farrugia:1991,Jones:2002,Savani:2011b} are commonly found within shock-driven sheath regions. \citet{Palmerio:2016} showed that these planar parts of the sheath in all likelihood are more geoeffective than non-planar parts. Due to enhanced solar wind dynamic pressure, sheath regions are capable of compressing the magnetosphere, pushing the dayside magnetopause down to below geosynchronous orbit \citep{Hietala:2014,Lugaz:2015b,Lugaz:2016,Kilpua:2019b}. The quintessential effect on the outer Van Allen radiation belts of such magnetospheric compression involves the drastic depletion of relativistic electron fluxes over a broad range of energy, equatorial pitch angle, and radial distance \citep{Pulkkinen:2007,Turner:2012,Hietala:2014,Hudson:2014,Kilpua:2015b,Alves:2016,Lugaz:2016,Xiang:2017}.

\justify

ME-driven sheath regions have only been studied for fast, shock-driving MEs and all these results about sheath properties and geoeffectiveness have been limited to sheaths associated with shocks. In addition, studies of such sheaths, associated with shocks, come mostly from simplified numerical simulations (often hydrodynamic) or similarities with Earth's bow shock. Though studies of ME-driven sheaths based on pure measurements have also been carried out, focusing on the generic profile and/or correlation/comparison with the driving ME \citep{Mitsakou:2009,Guo:2010,Mitsakou:2014,Kilpua:2017b,Jian:2018,Kilpua:2019b}, superposed epoch analysis \citep{Kilpua:2013,Masias:2016,Rodriguez:2016,Janvier:2019}, radial evolution \citep{Lugaz:2019,Good:2020}, to the best of our knowledge there has been no specific study of sheaths driven by MEs not driving any shocks. In this study, we examine ME-driven sheath regions measured by the twin STEREO \cite[Solar Terrestrial Relations Observatory;][]{Kaiser:2005} spacecraft from 2007-2016 focusing on all sheath regions, whether or not they are associated with shocks. As described above, we define the sheath as a region with a significant (see criteria in section~\ref{ssec:SDS}) increase in density and magnetic field strength for at least 2 hours above the background and directly preceding a ME. Through this study, we aim to examine the distinctness of MEs, especially the distinguishability of ME-driven sheaths with and without shocks.

\justify

The paper is organized as follows. In section~\ref{sec:methodology}, we explain the method of classifying distinct MEs by using the STEREO CME list of \citet{Jian:2018} and implementing an ``Automated Sheath Identification Algorithm" with visual inspection. In section~\ref{sec:Res}, we present our statistical results. A brief summary and discussion of our results are in section~\ref{sec:Dis}.

\section{Methodology} \label{sec:methodology}

\subsection{Initial Classification of MEs} \label{ssec:data} 

\justify

We classify the MEs measured by the twin STEREO spacecraft, as reported in \citet{Jian:2018} into three main categories. For each ME, \citet{Jian:2018} lists the start time of the CME, start time of the ME, and end time of the ME at the measuring spacecraft in this order. When the start time of the CME differs from the start time of the ME, it means there is a sheath region upstream of the ME. For MEs driving shocks, the CME start time corresponds to the arrival of the shock at the measuring spacecraft. The first category (Cat-I) includes MEs driving a shock near 1 AU and with a clear sheath region upstream of the ME. We use the Heliospheric shock database of \citet{Kilpua:2015a}, generated and maintained at the University of Helsinki (links for this and all other sources used in this study are listed in the acknowledgements section) and STEREO shock list for identifying shocks associated with these MEs. We refer to the former shock database as IPshocks database from now on. We only include shocks which are either listed in both of the aforementioned databases or only listed in one of them but have magnetosonic Mach numbers (M$_{ms}$) greater than 1. \textbf{For this category, we initially identify 105 MEs}. The second category (Cat-II) includes MEs with associated sheath regions but with no shock signatures or sometimes associated with a shock-like discontinuities. This category also includes MEs with shocks listed in the IPshocks database with M$_{ms}$\textless1. \textbf{We exclude 13 potential Cat-II MEs as the sheath boundaries for these MEs were not properly defined due to considerable data gaps. Therefore, initially through visual inspection, we identify 17 Cat-II MEs}. The third category (Cat-III) includes MEs driving no sheaths. For this specific category, the CME and ME start \textbf{times} listed in \citet{Jian:2018} \textbf{are} the same. \textbf{Initially, we identify 66 Cat-III MEs from the STEREO CME list}. We exclude 21 MEs through visual inspection as these MEs are associated with ``driverless shocks" or shocks which are not followed by their drivers \citep[see][]{Gopalswamy:2009}. These ``driverless shocks" are those for which the start time of the CME corresponds to a shock and is listed as the same time as the start of the ME, so that these CMEs have a shock and no sheath or more reasonably a shock, a sheath but no driver (ME).

\subsection{Automated Sheath Identification Algorithm} \label{ssec:SDS}

\justify

\textbf{In addition to the visual search described above}, we devise an Automated Sheath Identification Algorithm (ASIA) for the identification of any possible sheath region upstream of the ME. \textbf{The algorithm is used to provide a quantitative basis to confirm the initial classification and to reclassify as needed a few unclear events}. The sheath region in general features elevated levels of density and magnetic field strength over the background. In our approach, we approximate the sheath to be 1.5 times more compressed and 1.3 times more magnetized compared to the unperturbed solar wind. The rationale for these specific ratios is that the weakest shocks measured near 1 AU have a jump of $\sim$1.3 in the density and magnetic field strength at the shock but the density in the sheath is often more elevated than just downstream of the shock \citep{Manchester:2005}, which led to our criteria of 1.5 for the density and 1.3 for the magnetic field strength. Although enhancements in the solar wind speed, temperature, and dynamic pressure are also considered characteristic evidences for the identification of sheaths associated with shocks, such enhancements are not universally associated with sheaths not preceded by shocks. Therefore, for better optimization, we limit the ASIA to only search for predefined enhancements in density and magnetic field strength over the quiescent solar wind for the identification of sheaths.

\justify

\begin{figure*}[htbp!]
  \centering
        \includegraphics[width=1.0\linewidth]{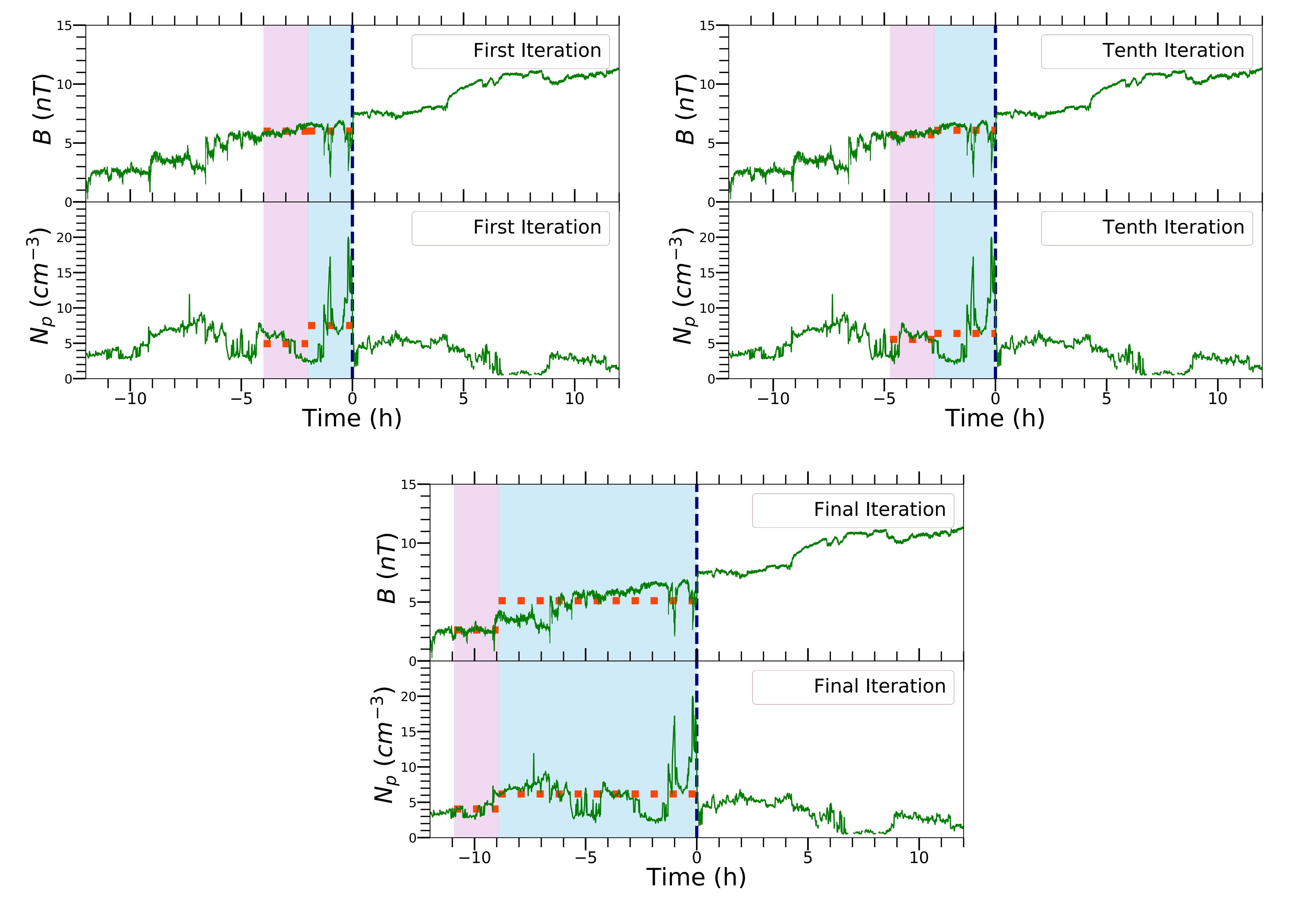}
        \caption{Visual representation (selected iterations) of the ASIA based on the average proton density and magnetic field strength criteria. The blue shaded region is the approximated sheath for each iteration and the pink shaded region is the approximated upstream solar wind. The horizontal red dotted lines in each region indicate the averages within that region. \textbf{The vertical navy dashed line marks the start of the ME leading edge.}}
         \label{fig:Detection}
  \end{figure*}

We initially start with a \textbf{preliminary} sheath, \textbf{extending up to the ME leading edge and with a duration of 2 hours} (blue shaded region in Figure~\ref{fig:Detection}) and we approximate the 2-hr interval prior to this \textbf{preliminary} sheath as the background solar wind (pink shaded region in Figure~\ref{fig:Detection}). The reason behind starting with this predefined sheath interval of 2 hours is that $\sim$97{\%} of the sheaths in our database are longer than 2 hours in duration. Therefore, we approximated a 2-hr interval to be the minimum threshold for any \textbf{preliminary} sheath. However, this results in the algorithm inherently not identifying the 4 sheaths in our database \textbf{that} are shorter than 2 hours. 

\justify

Then, for each following iteration of the ASIA, we extend this \textbf{preliminary} sheath by 5-min steps but always limit the background solar wind as a 2-hr interval prior to this. We carry out the iterations in a recursive manner and extend this \textbf{preliminary} sheath up to 30 hours before the start of the ME (\textbf{allowing identification} of extremely long sheaths) and each time compare the average density and magnetic field strength in this \textbf{preliminary} sheath with that of the background. As there can be substantial fluctuations of the solar wind plasma and interplanetary magnetic field (IMF) within the sheath, it is quite possible to have an interval that satisfies our average criteria, only for the region enclosed by the immediate next time step to violate one or both of them. To negate that, our goal is to find the time step after which for a prolonged interval, both of the ratios are satisfied. The interval enclosed by this specific time step upstream of the ME is identified as the probable sheath. However, as our ratios are somewhat subjective, we do not entirely rely on the ASIA for the identification of probable sheaths and no sheaths. For disagreements between the initial classification of MEs, solely based on the STEREO CME list of \citet{Jian:2018} and ASIA, we visually inspect the solar wind plasma and IMF profiles for each discrepancy. We run this algorithm for each and every event of our database. For the Cat-III events, the algorithm identified 20 events to have sheaths. We visually inspected these 20 events and identified them to have sheaths as well. We then looked for upstream shock signatures within the previous 30-hr period prior to the start of the ME in both of the shocks databases to determine whether to move these events to Cat-I or Cat-II. Based on the findings, we moved 1 event to Cat-I and 19 to Cat-II. In summary, the ASIA accurately identified 172 out of the 188 CMEs. Therefore, after the application of the ASIA along with thorough visual insepection, our final database is composed of 106 Cat-I, 36 Cat-II, and 46 Cat-III MEs.

\subsection{Example Events from Different Categories} \label{ssec:Ex}

\begin{figure*}[htbp!]
  \centering
        \includegraphics[width=1.0\linewidth]{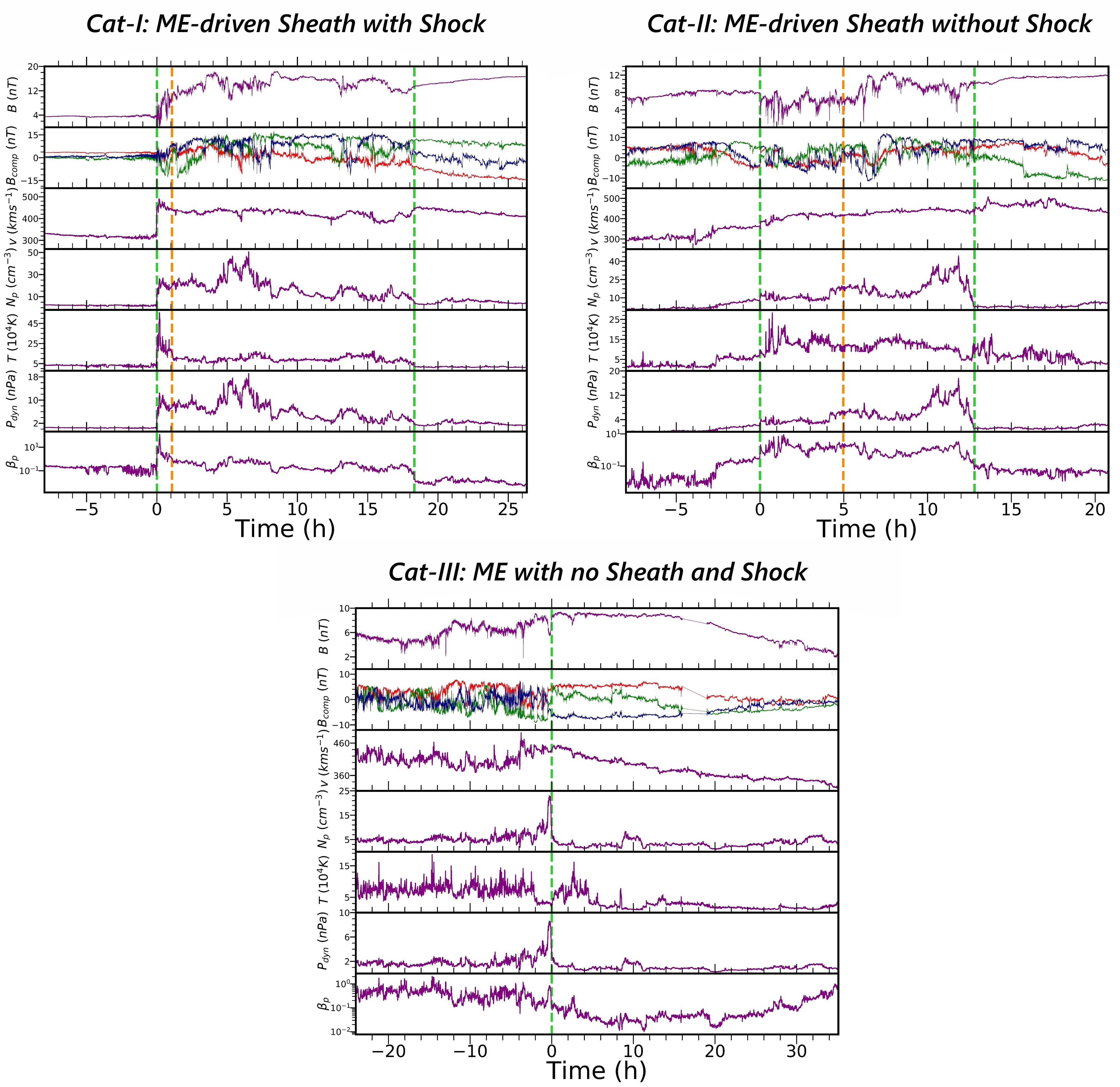}
        \caption{IMF and plasma measurements of three example CMEs. The seven panels show the magnetic field strength, components in RTN coordinates (red: radial, green: tangential, blue: normal), solar wind speed, proton density, proton temperature, dynamic pressure, and proton beta (from top to bottom). The two vertical green dashed lines bound the sheath, with t=0 corresponding to the start of the sheath and \textbf{the vertical orange dashed lines denote the start of the sheath, as identified by the ASIA (top panel)}. For the bottom panel, the start of the ME is denoted by the green dashed line, with t=0 corresponding to the start of the ME. \textit{Top left}: STEREO-B measurements of the Cat-I CME starting on 12 April 2014 at 2:27 UT, \textit{Top right}: STEREO-B measurements of the Cat-II CME starting on 29 October 2011 at 04:12 UT, \textit{Bottom}: STEREO-A measurements of the Cat-III CME starting on 28 April 2011 at 09:00 UT.}
         \label{fig:CS}
  \end{figure*}

\subsubsection{Cat-I (ME with an Associated Shock Structure and Sheath Region)} \label{sssec:One}

\justify

The Cat-I event presented here (see top left panel of Figure~\ref{fig:CS}) is driven by an expanding ME near 1 AU, with an expansion speed of 71 km\,s$^{-1}$. The expansion speed is taken as half the difference between the leading and trailing edge speeds \citep[see][]{Owens:2005}. The shock propagates in an upstream solar wind speed of 315 km\,s$^{-1}$. The upstream fast-magnetosonic speed is 63 km\,s$^{-1}$. Here, the upstream solar wind and fast-magnetosonic speeds are taken as the average of the 2-hr interval prior to the arrival of the shock at the spacecraft. The associated shock is listed in both the IPshocks database (M$_{ms}$ of 2) and the STEREO shock list (M$_{ms}$ of 2.6). The shock arrives at STEREO-B on 12 April 2014 at 2:27 UT. The spacecraft encounters the sheath for a relatively long period of 18.32 hours, with the leading edge of the ME arriving on 12 April 2014 at 20:46 UT. The ASIA also identified a sheath region upstream of the ME. For this particular event, the sheath interval identified by the ASIA is 17.25 hours, which is slightly shorter.

\subsubsection{Cat-II (ME with a Sheath but without a Shock)} \label{sssec:Two}

\justify

Our example Cat-II event reveals an interesting discontinuity. The start of this sheath coincides with a sharp decrease in the magnetic field strength (see top right panel of Figure~\ref{fig:CS}). The discontinuity, which probably is a slow mode shock/wave, arrives at STEREO-B on 29 October 2011 at 04:12 UT. According to the STEREO CME list of \citet{Jian:2018}, the spacecraft encounters the sheath for an interval of 7 hours. The ASIA also identifies a sheath region, extending out to 7.83 hours upstream of the ME. However, with visual inspection, we approximated the sheath to start at 12.80 hours before the start of the ME to accomodate the enhancements of the solar wind speed, proton density, temperature, and dynamic pressure. The driving ME expands at a much slower rate compared to the aforementioned Cat-I ME, with an expansion speed of $\sim$47 km\,s$^{-1}$. Here, the upstream solar wind speed of 357 km\,s$^{-1}$ might be an over-estimate because of the compression region right before the sheath. The upstream fast-magnetosonic speed is 72 km\,s$^{-1}$, which is slightly higher than in the previous example. The region upstream of the ME is a typical sheath-like structure as the physical parameters (proton density, temperature) within the sheath are enhanced by a factor of 2 than the background solar wind, while this factor is 3 for the dynamic pressure. The average speeds of the structures (sheath and ME) are also comparable to each other, indicating that the sheath is being driven by the ME.

\subsubsection{Cat-III (ME with No Associated Sheath Region)} \label{sssec:Three}

\justify

The example Cat-III ME presented in the bottom panel of Figure~\ref{fig:CS} arrives at STEREO-A on 28 April 2011 at 9:00 UT. The spacecraft encounters the ME for an interval of $\sim$35 hours before it ends on 29 April 2011 at 20:07 UT. The ME propagates in a moderately fast solar wind of 440 km\,s$^{-1}$. The upstream fast-magnetosonic speed is 55 km\,s$^{-1}$. This ME, with a leading edge speed of 441 km\,s$^{-1}$, is expanding near 1 AU with an expansion speed of $\sim$58 km\,s$^{-1}$. The ASIA identified no sheath region upstream of the ME. There are some notable enhancements and fluctuations in the magnetic field upstream of the ME. However, they do not coincide with any notable enhancements in the plasma parameters and might correspond to Alfv{\'e}n waves. The sharp increase in proton density just before the start of the ME, coinciding with a simultaneous sharp decrease in the magnetic field strength may be a product of magnetic reconnection. Also, the change in proton beta across the boundary is small, which is a condition favorable for reconnection \citep{Phan:2013}. This certainly is not an anomaly, as 8 out of the 46 Cat-III MEs exhibit such signatures. The ME has low average proton denisty (3.3 cm$^{-3}$) and temperature (2.7\e{4} K) compared to the background solar wind (11.6 cm$^{-3}$ and 3.2\e{4} K).

\section{Results} \label{sec:Res}

\justify

In this study, we use 1-min resolution plasma data from the PLASTIC instrument \citep{Galvin:2008} and 1/8 s high resolution magnetic field data from the IMPACT instrument \citep{Luhmann:2008} on-board the twin STEREO spacecraft.

\subsection{Proportion of Different Categories} \label{ssec:Proportion}

\begin{figure*}[htbp!]
  \centering
        \includegraphics[width=1.0\linewidth]{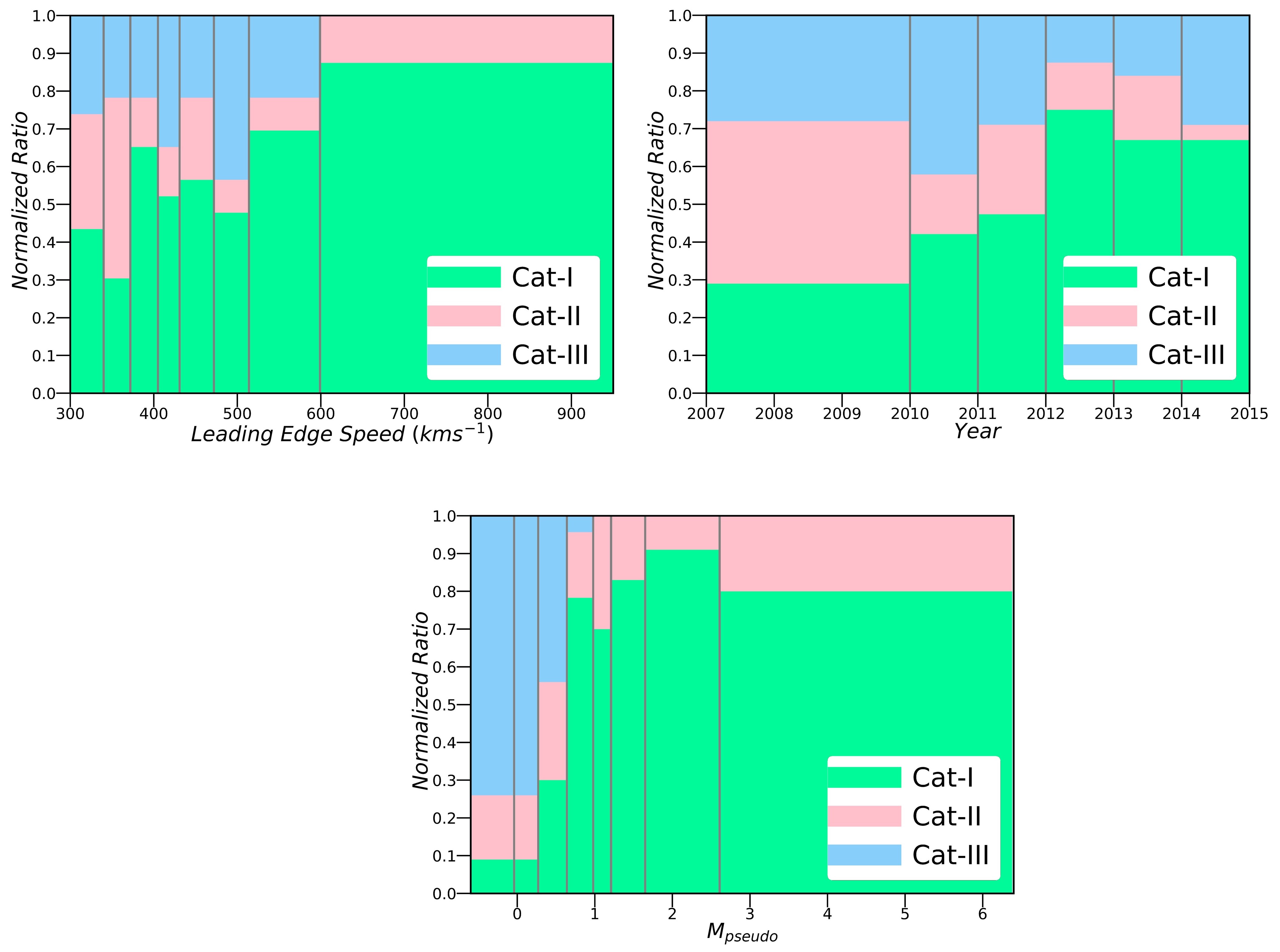}
        \caption{\textit{Top panel}: Normalized ratios of leading edge speeds (left) and annual occurrence rates (right) of the three ME categories, \textit{Bottom panel}: Normalized ratios of ``pseudo" Mach numbers (see section~\ref{sssec:magnetic ejecta}) of the three ME categories.}
         \label{fig:Proportion}
  \end{figure*}

\subsubsection{Variation in terms of the Leading Edge Speed of the ME} \label{sssec:LE}

\justify

We examine the proportion of different categories based on the leading edge speed of the ME near 1 AU (see top left panel of Figure~\ref{fig:Proportion}). We first arrange all the MEs by the increasing order of leading edge speeds. We exclude 3 MEs because the leading edge portions for these MEs have considerable data gaps. The leading edge speeds range from 301 km\,s$^{-1}$ (a Cat-III ME) to 949 km\,s$^{-1}$ (a Cat-I ME). Then, we bin the leading edge speeds into 8 intervals, each containing 23 MEs except the last one which contains 24. For each interval, we normalize the count from each category by the total number of MEs in that interval. We observe that, for small leading edge speeds, the number of MEs not driving shocks (Cat-II and Cat-III) outweigh the number of MEs driving shocks (Cat-I). For leading edge speeds below 400 km\,s$^{-1}$, only 42{\%} of MEs are able to drive shocks, confirming the results of the study by \citet{Lugaz:2017a} based on \textit{Wind}/ACE data. The normalized ratio of Cat-II MEs peak for leading edge speeds around 340-371 km\,s$^{-1}$, maybe indicating that they are in the process of forming shocks. As expected, with increasing leading edge speeds, the proportion of shock driving MEs \textbf{starts} outweighing the non-shock driving ones. However, the proportion of non-shock driving MEs does not immediately become negligible. Even for higher leading edge speed intervals, we do find \textbf{a} nonnegligible amount of non-shock driving MEs. For example, between leading edge speeds 590-710 km\,s$^{-1}$, 26{\%} of MEs still do not drive shocks. According to our database, MEs with leading edge speeds \textgreater 598 km\,s$^{-1}$ all drive sheaths and MEs with leading edge speeds \textgreater 701 km\,s$^{-1}$ all drive shocks near 1 AU and have associated sheaths. 

\subsubsection{Solar Cycle Variation} \label{sssec:SC}

\justify

In this section, we examine the annual variation of ME counts of the three categories. Different phases of two solar cycles are sampled: the solar minimum of solar cycle 23 and transition to solar cycle 24 (2007-2008), the rising phase of solar cycle 24 (2009-2011), and the double-peak solar maximum (2012-2014). For improved statistics, we combine the ME counts from 2007-2009. We then normalize the ME counts for each temporal bin (\textbf{2007-2009: 28, 2010: 19, 2011: 38, 2012: 40, 2013: 36, 2014: 27}, see \textbf{top right panel} of Figure~\ref{fig:Proportion}). We observe a gradual increase in the proportion of Cat-I MEs from 2009 onwards, coinciding with the rising phase of solar cycle 24 and peaking near the solar maximum (75{\%} of the MEs in 2012 are Cat-I). During solar maximum, a large proportion of MEs originate from active regions near big sunspot groups. Eruptive events associated with active regions usually give rise to fast MEs \citep{Manchester:2017}. Therefore, the increase in the occurrence rate of Cat-I MEs (fastest among the three categories) during the rising phase of a solar cycle is expected. In contrast, during solar minimum, a majority of MEs originate from streamer blowouts and quiescent prominences, located far from active regions. Such eruptions usually produce slow MEs \citep{Manchester:2017}. Therefore, as expected, the occurrence rate of Cat-II MEs (slowest among the three categories, \textbf{see Table~\ref{tab:Magnetic Ejecta} for comparison of average ME speeds of the three categories}) peak at solar minimum (33{\%} of Cat-II MEs occur during the solar minimum). The proportion of Cat-III MEs is relatively constant with a peak in 2010 during the rising phase of solar cycle 24. These MEs propagate in the fastest upstream solar wind and may be associated with slower MEs propagating through stream interaction regions (SIRs) that are more likely to occur outside of the solar maximum. 

\subsection{Superposed Epoch Analysis} \label{ssec:SEA}

\justify

We perform a superposed epoch analysis (SEA), also known as Chree analysis \citep{Chree:1913} to derive the generic profiles of sheaths driven by MEs near 1 AU. For Cat-I and Cat-II MEs, we use two characteristic epoch times, the start of the sheath and the start of the ME, so that all events with different timescales but from the same category are perfectly aligned. We primarily use the timings listed in \citet{Jian:2018}. Only for discrepancies between the STEREO CME list and the ASIA, we use the timings identified with visual inspection. For each event, the sheath and ME timescales are resampled to the average timescales of each category (average sheath duration: Cat-I is 9.27h, Cat-II is 8.32h). Then, the SEA is performed to statistically analyze the temporal profiles of Cat-I and Cat-II sheaths. This allows the average temporal profiles to be presented on the average timescales of the sheath and ME. We use the average sheath to ME interval ratio (0.44 for both Cat-I and Cat-II) to set up the typical timescale of the ME (typical ME duration: Cat-I is 21h, Cat-II is 18.9h). Therefore, the typical ME durations used in the SEA differ from the average ME durations listed in Table~\ref{tab:Magnetic Ejecta}. The sheaths of both categories are binned to 75 time bins and later averaged to get the same number of data points in the sheath interval. As the average ratio between the length of the sheath and ME intervals for both Cat-I and Cat-II MEs is 0.44, we impose 170 time bins for the ME (for both categories) and 16 (for Cat-I) or 18 (for Cat-II) time bins for the upstream solar wind (by comparing the average sheath duration of 9.27h for Cat-I and 8.32h for Cat-II with the predefined 2 hour interval upstream of the shock/discontinuity/sheath as the solar wind). Lastly, the averaged values from different events but of the same bin number are averaged to determine the average profiles of the physical parameters in the sheath and ME. 

\justify

For Cat-III MEs, we present the average timescale in a slightly different manner, as these MEs do not have associated sheath regions. We select \textbf{an extended} 8.8h solar wind interval upstream of the ME (average of the typical durations of Cat-I and Cat-II sheaths). Similar to Cat-I and Cat-II sheaths, we require this \textbf{extended pre-ME solar wind interval} to have 75 time bins. 211 time bins for the ME (by comparing the average ME duration of 24.8h with the 8.8h \textbf{extended pre-ME solar wind interval}) and 17 time bins for the solar wind \textbf{(2 hour interval upstream of the extended pre-ME solar wind interval)} are taken and later averaged.

\justify

\textbf{The goal of the extended 8.8h pre-ME solar wind is to identify, if it exists, any signature in the SEA of Cat-III MEs that may have been missed by the visual inspection. It is also used for visualization purposes to present all three categories with a similar format, the ME, preceded by a period of $\sim$8-9 hours and a 2 hour pristine solar wind}.

\begin{figure*}[htbp!]
  \centering
        \includegraphics[width=1.0\linewidth]{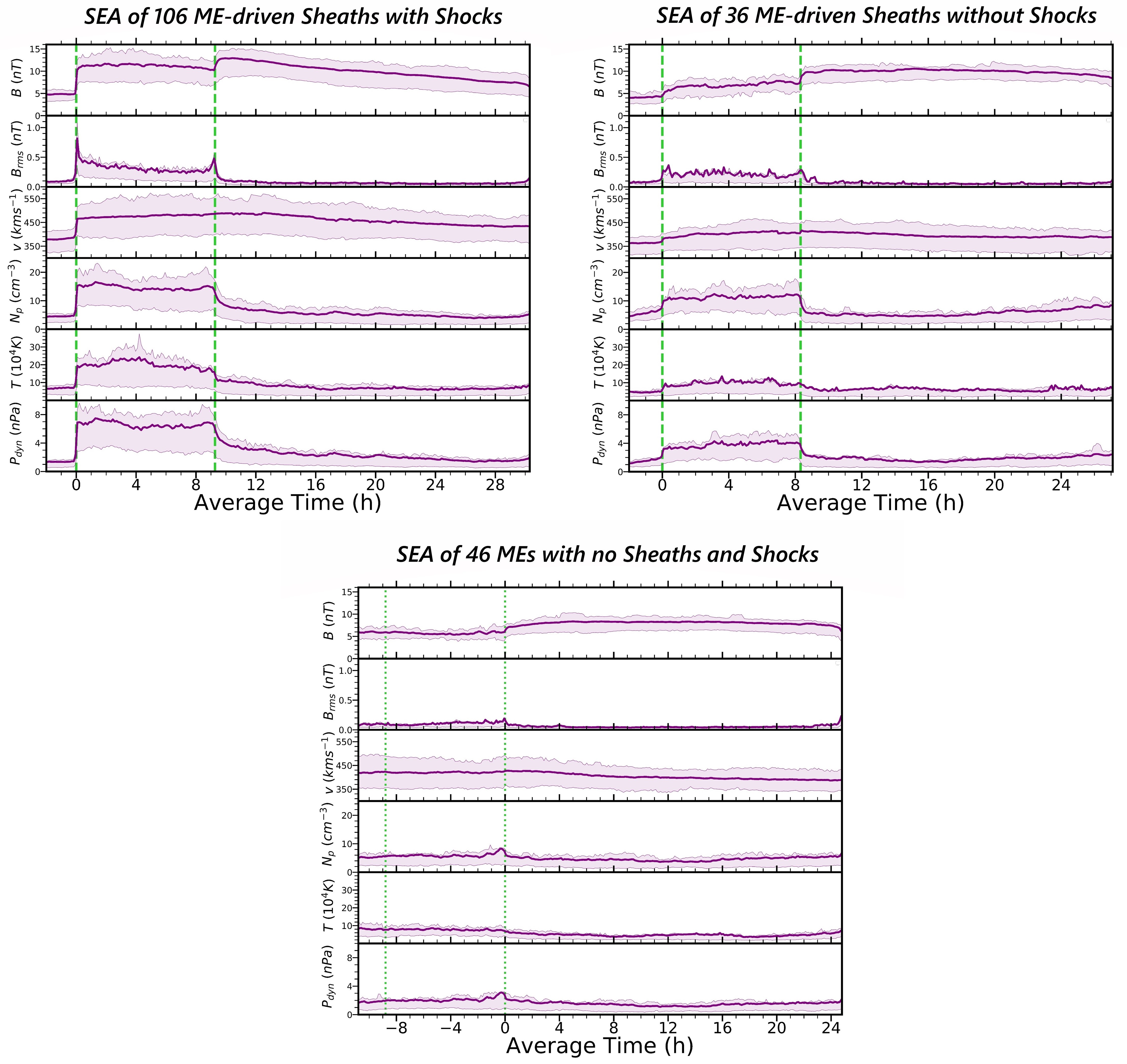}
        \caption{Superposed epoch analysis profiles for the 106 Cat-I CMEs (top left), 36 Cat-II CMEs (top right), and 46 Cat-III MEs (bottom). The purple curves show the average values and shaded regions indicate the interquartile ranges. The panels show distributions (from top to bottom) of the magnetic field strength, total root-mean-square error of the magnetic field strength, solar wind speed, proton density, proton temperature, and dynamic pressure. The vertical green dashed lines (top panel) bound the sheath region. The vertical green dotted lines (bottom panel) bound the \textbf{extended pre-ME solar wind interval}. The region to the left of the vertical green dashed lines represents the unperturbed solar wind and the region to the right of the vertical green dashed/dotted lines represents the entire ME.}
         \label{fig:SEA}
  \end{figure*}

\justify

From the SEA profiles of Cat-I CMEs (see top left panel of Figure~\ref{fig:SEA}), as expected, we observe an abrupt increase in the magnetic field strength, magnetic field fluctuations, proton density, speed, temperature, and dynamic pressure over the quiescent solar wind at the shock boundary. In the sheath region, the IMF and plasma profiles remain relatively constant (except the magnetic field fluctuations). We introduce a dynamic root-mean-square error of the magnetic field strength (B$_{rms}$) to quantify the magnetic field fluctuations. For each 1-min time step, we determine the fluctuations in the magnetic field strength within that time step by comparing the deviations of each value from the corresponding average of this 1-min time step. Therefore, the average B$_{rms}$ represents the arithmatic mean of the root-mean-square errors for each 1-min time step that makes up the solar wind/sheath/ME. The field fluctuations peak right at the shock arrival and then continuously decrease throughout the sheath, consistent with the study of \citet{Masias:2016}, before a second peak is observed at the leading edge of the ME. The piled-up IMF and plasma in the sheath correspond to enhanced magnetic field and plasma profiles over the background solar wind. The steady decline in the magnetic field strength and lower levels of proton density and temperature in the ME are direct consequences of the typical expansion undergone by the ME near 1 AU. In addition, the magnetic field strength inside the ME is asymmetric, owing to the aging effect \citep[see][]{Demoulin:2008}. The magnetic field fluctuations stabilise to lower typical solar wind levels in the ME, a feature characteristic of MCs \citep{Burlaga:1981}.

\justify

For Cat-II CMEs, at the start of the sheath, we also observe a jump in the parameters compared to the unperturbed solar wind (see top right panel of Figure~\ref{fig:SEA}). However, \textbf{the jumps are significantly less pronounced than for Cat-I CMEs}. In addition, the jumps are more extended and smoother, not as abrupt and sharp as Cat-I. Similar to Cat-I sheaths, the magnetic field and plasma profiles stay roughly constant within the sheath. The magnetic field fluctuations peak slightly after the start of the sheath and are more symmetric compared to Cat-I fluctuations. Cat-II sheaths also correspond to elevated IMF and plasma parameters over the background solar wind.

\justify

In the case of Cat-III CMEs (MEs), we observe an enhanced magnetic field in the ME compared to the ambient solar wind (see bottom panel of Figure~\ref{fig:SEA}). The magnetic field is roughly steady in the ME with relatively small fluctuations. The proton density and temperature are also low compared to the background solar wind.

\subsection{Average Profiles of the Structures} \label{ssec:Avg}

\subsubsection{Magnetic Ejecta Regions} \label{sssec:magnetic ejecta}

\justify

In Table~\ref{tab:Magnetic Ejecta}, we compare the average parameters within the ME for different categories. We also introduce three different types of Mach numbers. The ``pseudo" Mach number (M$_{pseudo}$) represents the ratio of the leading edge speed in the solar wind frame to the upstream fast-magnetosonic speed. The ``propagation" Mach number (M$_{propagation}$) denotes the ratio of the average ME speed in the solar wind frame to the upstream fast-magnetosonic speed. The ``expansion" Mach number (M$_{expansion}$) is the ratio of the expansion speed to the upstream fast-magnetosonic speed. The main results are as follows:

\justify

1) MEs driving Cat-I sheaths are the fastest among the three categories and they have the fastest expansion speed. 50{\%} of Cat-I MEs have an average speed in the solar wind frame which is less than the upstream fast-magnetosonic speed of 64 km\,s$^{-1}$.

\justify

2) MEs driving Cat-II sheaths are the slowest among the three categories and they have the slowest expansion speed as well. However, they also propagate in the slowest solar wind and that with the lowest fast-magnetosonic speed, i.e., in conditions that would be ideal to drive a shock. In fact, the average leading edge speed of MEs driving Cat-II sheaths is 416 km\,s$^{-1}$, which is almost identical to the minimum leading edge speed a Cat-II ME requires to drive a perpendicular shock near 1 AU, which is 412 km\,s$^{-1}$.

\justify

3) Cat-III MEs have almost the same average speed as those driving Cat-II sheaths. However, they propagate in the fastest solar wind and that with the highest fast-magnetosonic speed, i.e., in conditions that is the least ideal to drive a shock. The leading edge speeds of Cat-III MEs are comparable to the upstream solar wind speed, which is consistent with these MEs not driving any shocks. In addition, faster solar wind is less dense and also the average expansion speed of Cat-III MEs is moderate. As a result, it is easier for the deflected solar wind at the nose of the ME to go around it, which is a possible physical explanation as to why these MEs also do not drive sheaths.

\justify

4) Except for their faster speed, another major difference between MEs associated with Cat-I and Cat-II sheaths is that Cat-I MEs are of longer duration.

\justify

5) For Cat-I sheaths, the average magnetic field strength in the sheath (see Table~\ref{tab:Sheath}) is higher than that in the ME by 13{\%} and the peak magnetic field occurs 60{\%} of the time in the sheaths. For Cat-II sheaths, the magnetic field strength in the sheath is lower than that in the ME by 30{\%} and the peak magnetic field occurs only 25{\%} of the time in the sheaths. Cat-III MEs have the weakest average magnetic field strength.

\justify

6) While the leading edge speeds of the MEs organizes the three categories fairly well (see top left panel of Figure~\ref{fig:Proportion}), it is even better organized if we examine the distribution based on M$_{pseudo}$ (see \textbf{bottom} panel of Figure~\ref{fig:Proportion}), as defined above. We do the binning of M$_{pseudo}$ in a similar way to section~\ref{sssec:LE}. It is clearly evident that leading edge speeds of Cat-III MEs in the solar wind frame are considerably lower than the upstream fast-magnetosonic speed (M$_{pseudo}$\textless1). The largest proportion of Cat-II MEs are found in the range of 1\textless M$_{pseudo}$\textless 1.2, whereas $\sim$57{\%} of Cat-I MEs have M$_{pseudo}$\textgreater 1.2.

\begin{table}[htbp]
  \centering
  \caption{Averages and standard deviations of selected solar wind plasma and IMF parameters in the ME regions of different categories. The standard deviations represent 1-$\sigma$ uncertainty. The values in parentheses are of a subset of Cat-I MEs with leading edge speeds \textless 516 km\,s$^{-1}$.}
    \begin{tabular}{lllll}
    Parameter & Cat-I \textbf{(Cat-I Subset)} & Cat-II & Cat-III \\\hline\hline
    $<t>$ (h) & 27$\pm$16(27$\pm$18) & 22$\pm$10 & 25$\pm$13 \\\hline
    $<N_{p}>$ (cm$^{-3}$) & 5.4$\pm$3.7(6.0$\pm$3.8) & 5.9$\pm$4.2   & 4.6$\pm$3.3 \\\hline
    $<v>$ (km\,s$^{-1}$) & 460$\pm$112(402$\pm$57) & 396$\pm$88   & 402$\pm$65 \\\hline
    $<T>$ (10$^{4}$K) & 7.4$\pm$5.4(6.2$\pm$3.6) & 6.4$\pm$9.6 & 4.8$\pm$3.0 \\\hline
    $<B>$ (nT) & 9.9$\pm$3.9(9.5$\pm$3.1) & 9.9$\pm$2.9   & 8.0$\pm$3.1 \\\hline
    $<P_{dyn}>$ (nPa) & 2.2$\pm$1.7(2.0$\pm$1.4) & 1.9$\pm$2.1   & 1.5$\pm$1.1 \\\hline
    $<v>_{Upstream-sw}$ (km\,s$^{-1}$) & 382$\pm$81(345$\pm$44) & 364$\pm$68   & 421$\pm$82 \\\hline
    $<v>_{Upstream-fms}$ (km\,s$^{-1}$) & 64$\pm$26(53$\pm$18) & 48$\pm$18    & 73$\pm$35 \\\hline
    $v_{LE}$ (km\,s$^{-1}$) & 483$\pm$118(416$\pm$56) & 416$\pm$100   & 429$\pm$80 \\\hline
    $v_{exp}$ (km\,s$^{-1}$) & 26$\pm$38(12$\pm$26) & 12$\pm$26 & 17$\pm$30 \\\hline
    $M_{pseudo}$ & 1.7$\pm$1.2(1.5$\pm$1.1) & 1.1$\pm$1.1 & 0.1$\pm$0.3 \\\hline
    $M_{propagation}$ & 1.4$\pm$1.2(1.2$\pm$1.1) & 0.7$\pm$0.9 & -0.3$\pm$0.6 \\\hline
    $M_{expansion}$ & 0.4$\pm$0.6(0.2$\pm$0.5) & 0.3$\pm$0.5 & 0.3$\pm$0.6 \\\hline
    $<\beta>_{Proton}$ & 0.2$\pm$0.2(0.2$\pm$0.2) & 0.1$\pm$0.1 & 0.1$\pm$0.1 \\\hline
    \end{tabular}%
  \label{tab:Magnetic Ejecta}%
\end{table}%

\subsubsection{Sheath Regions} \label{sssec:Sheath}

\justify

We list the average sheath parameters driven by Cat-I and Cat-II MEs in Table~\ref{tab:Sheath}. Table~\ref{tab:t-test} lists the results from the Welch's t-test (with 95{\%} confidence level) to determine if differences between Cat-I and Cat-II sheaths are statistically different. In this section, we focus on the main differences between Cat-I and Cat-II sheaths, listing only those which are statistically significant. We find the following results:

\begin{table}[htbp]
  \centering
  \caption{Averages and standard deviations of selected solar wind plasma and IMF parameters in the sheath regions of different categories. The standard deviations represent 1-$\sigma$ uncertainty. The values in parentheses are of a subset of Cat-I sheaths that are driven by MEs with leading edge speeds \textless 516 km\,s$^{-1}$. $<Ratio>$: Ratio of the average values within the sheath to that in the upstream solar wind.}
    \begin{tabular}{lll}
    Parameter & Cat-I (Cat-I Subset) & Cat-II \\\hline\hline
    $<t>$ (h) & 9.3$\pm$4.9(9.3$\pm$5.2) & 8.3$\pm$4.0 \\\hline
    $<N_{p}>$ (cm$^{-3}$) & 14.4$\pm$8.0(15.8$\pm$8.2) & 11.3$\pm$6.1 \\\hline
    $<v>$ (km\,s$^{-1}$) & 477$\pm$113(411$\pm$57) & 404$\pm$91 \\\hline
    $<T>$ (10$^{5}$K) & 2.0$\pm$1.6(1.2$\pm$0.9) & 1.0$\pm$1.0 \\\hline
    $<B>$ (nT) & 11.2$\pm$4.6(9.8$\pm$3.3) & 6.9$\pm$2.6 \\\hline
    $<B_{rms}>$ (nT) & 0.3$\pm$0.2(0.3$\pm$0.2) & 0.2$\pm$0.4 \\\hline
    $<P_{dyn}>$ (nPa) & 6.5$\pm$4.4(5.4$\pm$3.0) & 3.8$\pm$2.6 \\\hline
    $<Ratio>_{N_{p}}$ & 3.8$\pm$2.3(4.0$\pm$2.1) & 2.2$\pm$1.4 \\\hline
    $<Ratio>_{v}$ & 1.3$\pm$0.2(1.2$\pm$0.1) & 1.1$\pm$0.1 \\\hline
    $<Ratio>_{T}$ & 3.4$\pm$2.6(3.1$\pm$2.4) & 2.1$\pm$1.5 \\\hline
    $<Ratio>_{B}$ & 2.6$\pm$1.4(2.7$\pm$1.5) & 1.8$\pm$0.9 \\\hline
    $<Ratio>_{P_{dyn}}$ & 6.4$\pm$4.7(6.0$\pm$4.2) & 2.8$\pm$2.0 \\\hline
    \end{tabular}%
  \label{tab:Sheath}%
\end{table}%

\justify

1) The compression in density and magnetic field in the sheath as compared to the background solar wind is significantly higher for Cat-I sheaths than Cat-II sheaths. The level of compression ($\sim$4) in density for Cat-I sheaths is unlikely to be explained only by the compression resulting from the associated shock, since shock compression ratio for a typical ME-driven shock is $\sim$2. This indicates that there is follow-up compression after the shock, consistent with the numerical simulation of \citet{Manchester:2005}.

\justify

2) Cat-I sheaths are made of denser (by 27{\%}), hotter (by 100{\%}), and more magnetized (by 62{\%}) plasma than Cat-II sheaths. The sheath plasma is faster (by 18{\%})  and has a higher dynamic pressure (by 71{\%}) in Cat-I than Cat-II sheaths. The level of fluctuations in the magnetic field is however not statistically different.

\justify

3) The average sheath durations for Cat-I and Cat-II sheaths are similar, at $\sim$8-9 hours. Because Cat-I sheaths are faster, their sizes are statistically larger than Cat-II sheaths.

\begin{table}[htbp]
  \centering
  \caption{Measure of statistical difference with 95{\%} confidence level (p-values from the Welch's t-test) between selected solar wind plasma and IMF parameters. P-values representing statistical difference are made bold.}
    \begin{tabular}{lllll}
    Parameter & {Cat-I \& Cat-II} & {Cat-I Subset \& Cat-II} \\\hline\hline
    $<Size>_{Sheath}$ & \textbf{0.0020} & 0.1253 \\\hline
    $<B>_{Sheath}$ & \textbf{4.2E-10} & \textbf{3.1E-06} \\\hline
    $<N_{p}>_{Sheath}$ & \textbf{0.0134} & \textbf{0.0021} \\\hline
    $<v>_{Sheath}$ & \textbf{0.0003} & 0.6806 \\\hline
    $<T>_{Sheath}$ & \textbf{2.8E-05} & 0.2126 \\\hline
    $<P_{dyn}>_{Sheath}$ & \textbf{2.1E-05} & \textbf{0.0067} \\\hline
    $<B_{rms}>_{Sheath}$ & 0.1780 & 0.6056 \\\hline
    $<v>_{Upstream-sw}$ & 0.2190 & 0.1326 \\\hline
    $<v>_{Upstream-fms}$ & \textbf{0.0002} & 0.1724 \\\hline
    $v_{LE}$ & \textbf{0.0016} & 0.9976 \\\hline
    $v_{exp}$ & \textbf{0.0156} & 0.9134 \\\hline
    $M_{pseudo}$ & \textbf{0.0065} & 0.0952 \\\hline
    $M_{propagation}$ & \textbf{0.0007} & \textbf{0.0060} \\\hline
    $M_{expansion}$ & 0.3111 & 0.8011 \\\hline
    \end{tabular}%
  \label{tab:t-test}%
\end{table}%

\subsubsection{Comparison of Cat-II Sheaths and Cat-I Sheaths Driven by Slower MEs} \label{sssec:Comp}

\justify

We next investigate if the differences between Cat-I and Cat-II sheaths are due to the different speeds of the drivers or they are inherently different. To do so, we create a subset of Cat-I MEs, where the leading edge speeds are \textless 516 km\,s$^{-1}$ (average leading edge speed of Cat-II MEs is 416$\pm$100 km\,s$^{-1}$). We list the average values of this subset in parentheses in the Cat-I column in Table~\ref{tab:Magnetic Ejecta} and Table~\ref{tab:Sheath}. From the comparison of upstream conditions and ME speeds, we can approximate that this subset of Cat-I sheaths and Cat-II sheaths are driven by similar MEs, propagating in similar upstream solar wind (see Table~\ref{tab:t-test}). We do see that the average properties of Cat-II sheaths and sheaths of this subset are more similar to each other than Cat-I and Cat-II sheaths. This is true for the sheath size, speed, and temperature (see Table~\ref{tab:t-test}). However, the sheath density, magnetic field strength, and dynamic pressure remain \textbf{more} statistically elevated in Cat-I sheaths driven by slower MEs than in Cat-II sheaths (see Table~\ref{tab:Sheath} and Table~\ref{tab:t-test}). This highlights intrinsic differences in the compression (of the density and magnetic field) associated with the presence or absence of a shock.

\justify

We expect the comparison between this subset of Cat-I sheaths and Cat-II sheaths to be statistically less significant, as they are driven by MEs with similar speeds. Except the sheath density and the upstream solar wind speed, we indeed observe higher p-values for all other examined parameters (see Table~\ref{tab:t-test}). From these p-values, we can approximate that this subset of Cat-I sheaths and Cat-II sheaths are similar in many aspects, even though their formation mechanisms are supposed to be different, as Cat-I sheaths are driven by MEs which also drive shocks and Cat-II sheaths are \textbf{associated} with MEs that do not drive shocks.

\subsection{Correlation Analysis between the Sheath and ME} \label{ssec:CorAnalysis}

\justify

We investigate any potential correlation between the average parameters of the structures. We apply a linear least-squares regression technique to test the correlation of the average parameters (proton density, solar wind speed, and magnetic field strength) within the sheath with that of the driving ME. We use the average ME speed in our correlation analysis to exclude the contribution of the expansion speed, similar to \citet{Gopalswamy:2008}. In the STEREO CME list, \citet{Jian:2018} ranks the MEs with a quality index ranging from 2 to 0. MEs with an index of 2 are the closest to the MC criteria \citep[enhanced magnetic field strength, smooth rotation of the magnetic field vector, and low proton temperature, see][]{Burlaga:1981}. MEs with a quality index 1 are MC-like, with the quality index 0 corresponding to non-MCs. Therefore, for each category, we use two sets to examine the correlation of the sheath with its driver ME. The first set considers all ME types \textbf{(with quality index 0, 1, and 2)} and the second set considers MEs with quality index 1 and 2 (see Table~\ref{tab:Corr}).

\subsubsection{Correlation between Average Sheath and ME Properties} \label{sssec:CR}

\justify

\begin{figure*}[htbp!]
  \centering
        \includegraphics[width=0.9\linewidth]{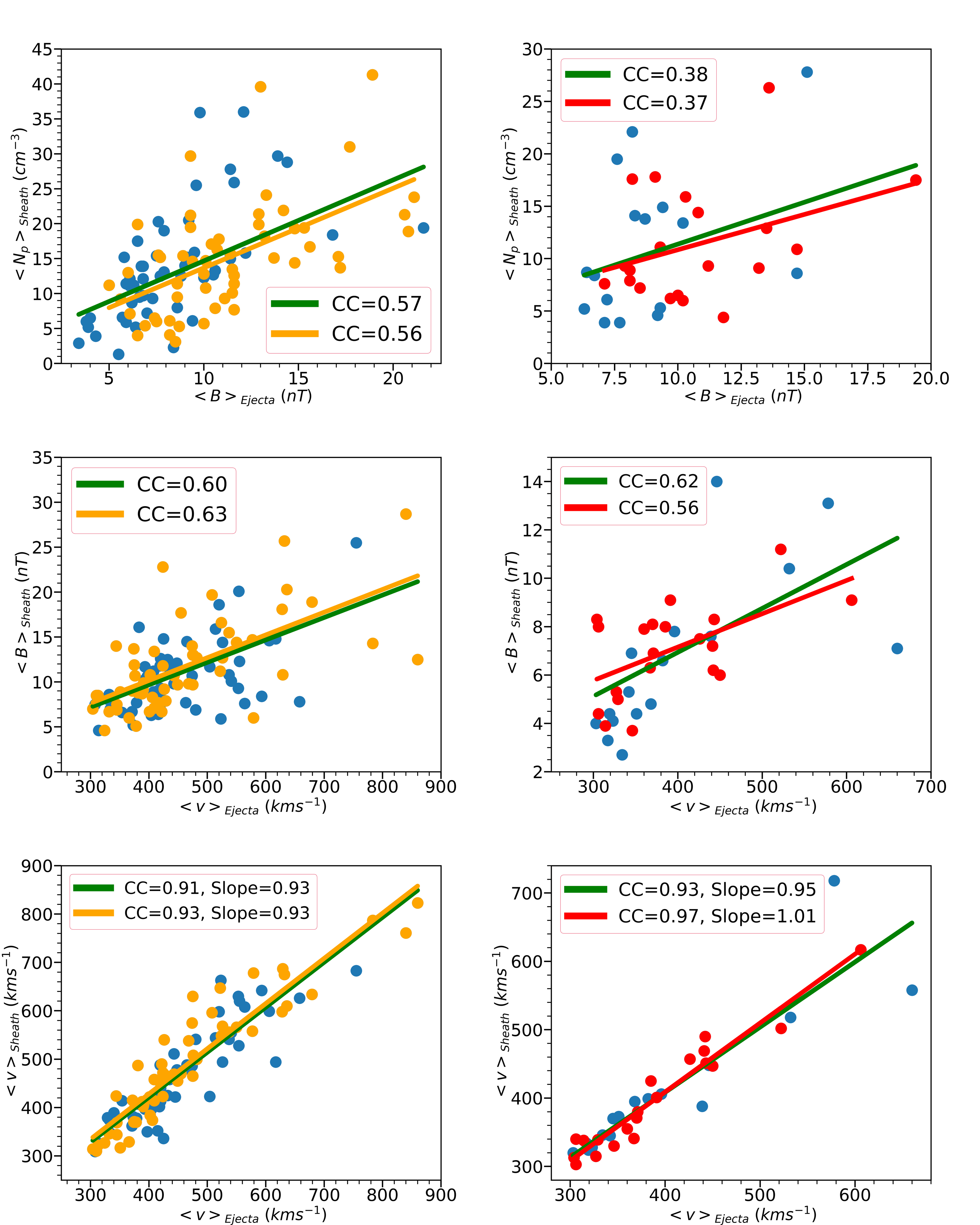}
        \caption{Scatter plots of selected average parameters for Cat-I (left panels) and Cat-II (right panels). Blue circles represent observations of MEs \textbf{with quality index 0}. Orange (for Cat-I) and red (for Cat-II) circles represent observations of MEs \textbf{with quality index 1 and 2}. The solid green line represents the best linear least-squares fit of the data set for all ME observations \textbf{(with quality index 0, 1, and 2)}. The orange (for Cat-I) and red (for Cat-II) lines represent the same but only for observations of MEs \textbf{with quality index 1 and 2}.}
         \label{fig:ICor}
  \end{figure*}

1) The average proton density of Cat-I sheaths show modest (Correlation Coefficient or CC=0.50 - 0.75) positive correlation with the average proton density (CC=0.55) and magnetic field strength (CC=0.57, see Figure~\ref{fig:ICor}) of the ME, which mean that sheaths driven by denser and stronger MEs tend to be denser as well. However, the average proton density of Cat-I sheaths does not have any sort of relationship with how fast the ME is propagating. This is surprising as the rate at which the ME accumulates solar wind material over its propagation is assumed to be dependent on the speed of the ME. On the other hand, density and speed of general solar wind are typically anti-correlated, i.e., fast solar wind with lower density while slow solar wind with higher density. Thus, the two effects may result in no clear correlation between the sheath density and ME speed. For Cat-II sheaths, the average proton density does not have any clear correlation with the average parameters of the ME (see Figure~\ref{fig:ICor} for the correlation with the average magnetic field strength of the ME).

\begin{table}[htbp]
  \centering
  \caption{Correlation coefficients of the linear least-squares regression between the average parameters of the structures \textbf{(sheath and ME)} for different categories. The values in parentheses refer to linear least-squares regression including MEs \textbf{with quality index 1 and 2} only. + indicates positive correlation with CC\textless0.25. - indicates negative correlation with CC\textless0.25. CCs\textgreater0.50 are made bold.}
    \begin{tabular}{lllllll}
    \multicolumn{1}{p{6em}}{Parameter} & \multicolumn{1}{p{4em}}{Average N$_{p,Ejecta}$} & \multicolumn{1}{p{4em}}{Average N$_{p,Ejecta}$} & \multicolumn{1}{p{4em}}{Average B$_{Ejecta}$} & \multicolumn{1}{p{4em}}{Average B$_{Ejecta}$} & \multicolumn{1}{p{4em}}{Average v$_{Ejecta}$} & \multicolumn{1}{p{4em}}{Average v$_{Ejecta}$} \\
          & \multicolumn{1}{p{4em}}{Cat-I} & \multicolumn{1}{p{4em}}{Cat-II} & \multicolumn{1}{p{4em}}{Cat-I} & \multicolumn{1}{p{4em}}{Cat-II} & \multicolumn{1}{p{4em}}{Cat-I} & \multicolumn{1}{p{4em}}{Cat-II} \\\\\hline\hline
    $<N_{p}>_{Sheath}$ & \textbf{0.55} (\textbf{0.52}) & + (+) & \textbf{0.57} (\textbf{0.56}) & 0.38 (0.37) & - (-) & - (-) \\\hline
    $<B>_{Sheath}$ & + (-) & + (-0.25) & \textbf{0.64} (\textbf{0.62}) & \textbf{0.54} (0.34) & \textbf{0.60} (\textbf{0.63}) & \textbf{0.62} (\textbf{0.56}) \\\hline
    $<v>_{Sheath}$ & -0.30 (-0.36) & + (-0.32) & + (+) & 0.31 (+) & \textbf{0.91} (\textbf{0.93}) & \textbf{0.93} (\textbf{0.97}) \\\hline
    \end{tabular}%
  \label{tab:Corr}%
\end{table}%

\justify

2) The average magnetic field strength of Cat-I sheaths also shows a modest relationship with the magnetic field strength (CC=0.64) and the average speed (CC=0.60, see Figure~\ref{fig:ICor}) of the ME. Same is true for Cat-II sheaths. Sheaths of both categories do not have any substantial correlation with the average proton density of the ME. 

\justify

3) The average speeds of Cat-I and Cat-II sheaths do not have any clear correlation with the average proton density and magnetic field strength of the ME. The average speeds in the sheath for both categories only depend upon how fast the driving ME is propagating (CC=0.91 for Cat-I, CC=0.93 for Cat-II, see Figure~\ref{fig:ICor}), which means, as expected, a faster ME drives a faster sheath as well.

\subsubsection{Correlation of the Sheath Size with ME Properties} \label{sssec:Size}

\justify

\begin{figure*}[htbp!]
  \centering
        \includegraphics[width=0.9\linewidth]{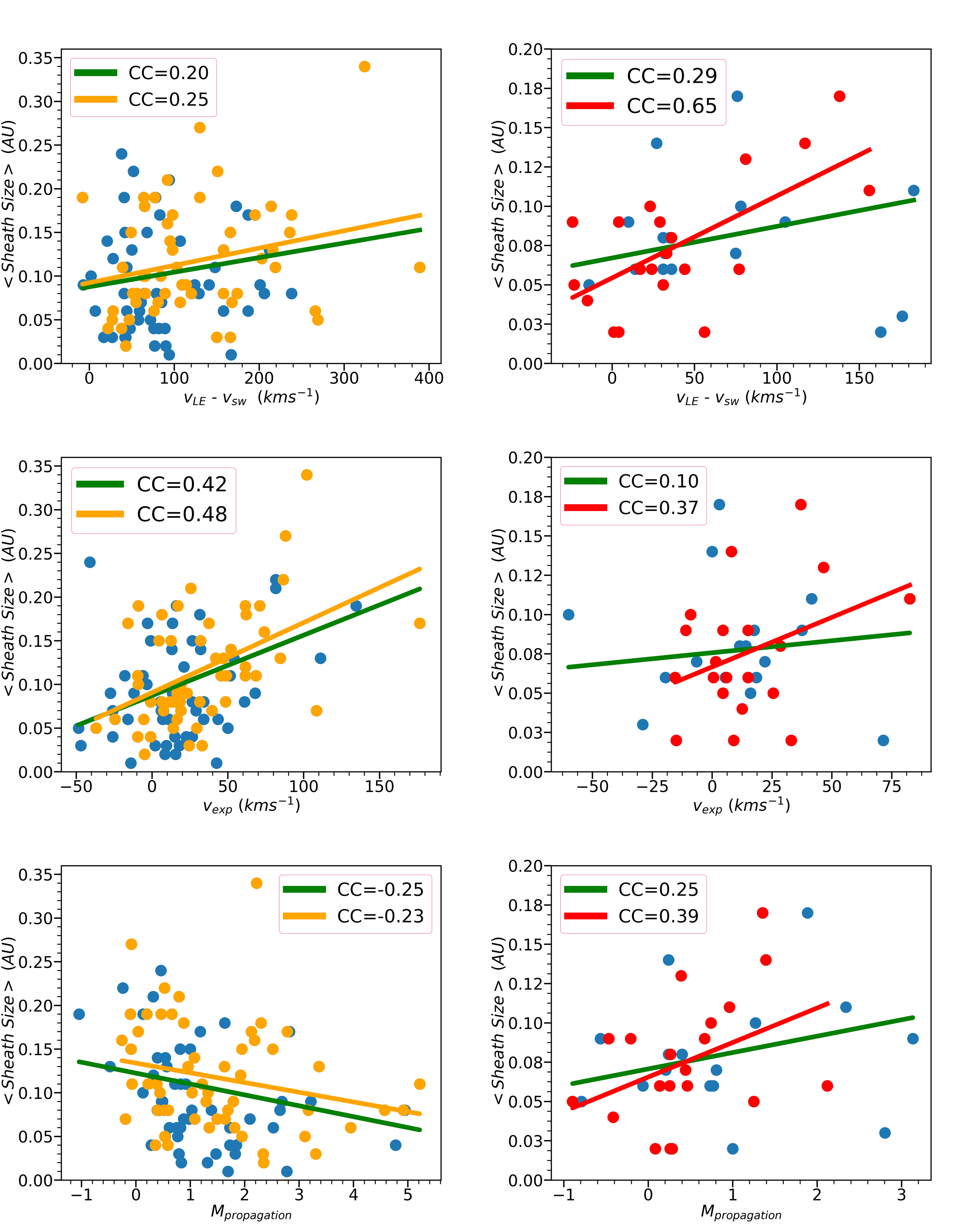}
        \caption{Scatter plots of the sheath size in radial direction (in AU) and selected ME parameters for Cat-I (left panels) and Cat-II (right panels). Blue circles represent observations of MEs \textbf{with quality index 0}. Orange (for Cat-I) and red (for Cat-II) circles represent observations of MEs \textbf{with quality index 1 and 2}. The solid green line represents the best linear least-squares fit of the data set for all ME observations \textbf{(with quality index 0, 1, and 2)}. The orange (for Cat-I) and red (for Cat-II) lines represent the same but only for observations of MEs \textbf{with quality index 1 and 2}.}
         \label{fig:SCor}
  \end{figure*}

We also examined the linear relationship of the sheath size (in AU) of Cat-I and Cat-II sheaths with the speed profile and associated Mach numbers of the ME (see Table~\ref{tab:Corsize} and Figure~\ref{fig:SCor}). The sheath size is approximated as the product of the average speed in the radial direction and the time interval through which the spacecraft measures the sheath. \textbf{Here, this approximation for the sheath size is dependent on the location where the sheath is crossed with respect to the nose of the ME}.

\begin{table}[htbp]
  \centering
  \caption{Correlation coefficients of the linear least-squares regression between the sheath size in radial direction (AU) with specific ME parameters for different categories and different combinations. (+) indicates positive correlation with CC\textless0.25, (-) indicates negative correlation with CC\textless0.25. CCs\textgreater0.50 are made bold.}
    \begin{tabular}{p{16em}p{3em}cc}
    \multicolumn{1}{c}{Parameter} & \multicolumn{3}{c}{Sheath Size (AU)} \\\hline\hline
    \multicolumn{1}{c}{} & \multicolumn{1}{c}{Category} & All MEs & MEs (Quality Index 1 \& 2) \\
    $v_{LE}$ (km\,s$^{-1}$) & Cat-I & 0.39  & 0.44 \\\hline
    $v_{LE}$ (km\,s$^{-1}$) & Cat-II & (+) & 0.38 \\\hline
    $v_{LE}$ - $<v>_{Upstream-sw}$ (km\,s$^{-1}$) & Cat-I & (+)  & 0.25 \\\hline
    $v_{LE}$ - $<v>_{Upstream-sw}$ (km\,s$^{-1}$) & Cat-II & 0.29  & \textbf{0.65} \\\hline
    $<v>_{Ejecta}$ (km\,s$^{-1}$) & Cat-I & (+)  & 0.28 \\\hline
    $<v>_{Ejecta}$ (km\,s$^{-1}$) & Cat-II & (+)  & 0.33 \\\hline
    $v_{exp}$ (km\,s$^{-1}$) & Cat-I & 0.42  & 0.48 \\\hline
    $v_{exp}$ (km\,s$^{-1}$) & Cat-II & (+)  & 0.37 \\\hline
    $M_{pseudo}$   & Cat-I & (-) & (-) \\\hline
    $M_{pseudo}$   & Cat-II & (+)  & 0.41 \\\hline
    $M_{propagation}$ & Cat-I & -0.25 & (-) \\\hline
    $M_{propagation}$ & Cat-II & 0.25  & 0.39 \\\hline
    $M_{expansion}$ & Cat-I & 0.30  & 0.35 \\\hline
    $M_{expansion}$ & Cat-II & (-) & 0.25 \\\hline
    \end{tabular}%
  \label{tab:Corsize}%
\end{table}%

\justify

The sheath size for both Cat-I and Cat-II sheaths show weak linear correlation with the speed profile of the driving ME. The best correlations for Cat-I MEs are found with the leading edge speed and the expansion speed ($\sim$0.4). In addition, both these correlations are positive, indicating that faster CMEs have thicker sheaths. The correlations in general are poorer for Cat-II MEs. There exist similar poor correlations between the sheath size and the three aforementioned Mach numbers for both categories.

\justify

This poor correlation between the sheath size and the speed profile and associated Mach numbers of the driving ME is unexpected compared to the results of \citet{Russell:2002,Savani:2011b}, where sheath thickness is approximated to decrease with the shock Mach number, meaning slower MEs tend to have thicker sheaths. On a similar note, \citet{Masias:2016} found slower MCs to have thicker sheaths on average at 1 AU owing to the effect of drag during their propagation. 

\section{\textbf{Discussion and Conclusions}} \label{sec:Dis}

\justify

In this manuscript, we have outlined solar wind plasma and IMF properties in ME-driven sheath regions. We examined 142 sheaths and 188 MEs measured by the twin STEREO spacecraft near 1 AU during 2007-2016. We classified these MEs into three categories, based on whether a ME drives a shock and a sheath, or only a sheath, or neither. We determined the statistical differences between sheaths, with and without shocks. We also investigated the correlation between the average properties of the sheath and ME. 

\justify

As expected, MEs driving both shocks and sheaths are the fastest MEs \citep[see][]{Gopalswamy:2006}, propagating through typical solar wind with moderately high magnetosonic speed. MEs driving only sheaths are the slowest MEs, propagating through slower solar wind with low magnetosonic speed. MEs without sheaths and shocks are also slower MEs, propagating through faster solar wind with high magnetosonic speed. 

\justify

According to our findings, $\sim$56{\%} of MEs drive shocks near 1 AU. This shock rate varies in phase with solar cycle activity \citep[e.g.,][]{Lindsay:1994,Gopalswamy:2003,Jian:2011,Jian:2018}. Significant proportion (92{\%}) of these faster MEs (who drive shocks near 1 AU) occur during the rising and maximum phase of the solar cycle. In contrast, MEs that propagate with a speed slightly higher than the ambient solar wind and expand slowly drive sheaths, even if they are not fast enough to drive shocks. $\sim$43{\%} of the MEs that occur during the solar minimum are these slowest MEs. Their sheaths are about as long as a typical sheath and have some similarities in properties with sheaths preceded by shocks, once correcting for the leading edge speed of the ME. Occurrence of MEs that do not drive sheaths is relatively constant through the solar cycle and is slightly higher in the rising phase

\justify

In our database, we find that a ME with a leading edge speed as low as 316 km\,s$^{-1}$ drives a shock near 1 AU, whereas another one with a leading edge speed 701 km\,s$^{-1}$ does not. Such scenarios were also reported by \citet{Chi:2016}. According to them, these instances can be either subjected to the variation of the background solar wind plasma and IMF conditions for individual cases \citep[e.g.,][]{Shen:2007} or to the effect of drag during the propagation of MEs in interplanetary space \citep[e.g.,][]{Lugaz:2013a,Vrsnak:2013}. Confirming \citet{Lugaz:2017a}, we find that MEs need to have a leading edge speed faster than 515 km\,s$^{-1}$ to have clearly a \textgreater 60{\%} probability of driving both a shock and a sheath near 1 AU.

\justify

From our statistical analysis, we find that ME-driven sheaths are significantly more compressed and magnetized with higher dynamic pressure compared to the quiescent ambient solar wind. Sheaths preceded by shocks are on average hotter, faster, compressed, and magnetized with higher magnetic field fluctuations and dynamic pressure compared to sheaths without shocks. In addition, shock driving MEs are more magnetized than non-shock driving MEs \citep[see][]{Wu:2016}. Some of these differences in the sheath regions become smaller when sheaths driven by MEs (with and without shocks) with similar speed profiles are compared, though even in this case the average proton density, magnetic field strength, and dynamic pressure still remain statistically higher for sheaths with shocks than sheaths without shocks. These properties are some of the most important ones to determine the effect of solar wind structures on geospace, both in terms of reconnection and compression. As such, we expect sheaths preceded by shocks to have a stronger effect on the radiation belt and magnetosphere for a given speed than sheaths not preceded by shocks.

\justify

From the investigated parameters, we find a strong correlation between the average speed of the structures \textbf{(sheath and ME)} for any category of ME-driven sheaths (with and without shocks), consistent with previous studies \citep{Mitsakou:2009,Janvier:2019}. Such strong correlations confirm that sheaths are indeed driven by these MEs, even in the absence of shocks. \citet{Kilpua:2019b} found similar strong correlation between the average speed within the sheath and the leading edge speed of the ME. The sheath magnetic field strength is also moderately correlated with the average magnetic field strength \citep[see][]{Chi:2016} and speed of the ME for all sheaths (with and without shocks). In the case of the average proton density in the sheath, it has moderate correlation with the average magnetic field strength of the ME for sheaths with shocks. There is no correlation between these parameters for sheaths without shocks.

\justify

The lack of correlation we find between the radial thickness of sheaths and the speed profile and associated Mach numbers of the driving ME, irrespective of the type of sheath is surprising. However, it is in perfect agreement with \citet{Salman:2020} where we found that typical sheath durations throughout the inner heliosphere are independent of the initial ME speed. This finding runs contrary to the conventional wisdom of how these sheaths form and may indicate that accumulation of the solar wind, before the ME drives a shock is the dominant mechanism that forms sheaths near 1 AU, rather than the shock compression, since for shock compression, the stand-off distance (the radial separation between the magnetic obstacle boundary and the shock front) should be related to the ME speed or the shock Mach number \citep{Russell:2002,Savani:2011b}. Important to note that, the time period concerning this study primarily covers the significantly weak solar cycle 24 compared to solar cycle 23 \citep{McComas:2013} where the fraction of fast CMEs is not dominantly high \citep[see Figures 7 and 18 in ][]{Jian:2018}. Further studies, extending this for example with ACE or \textit{Wind} data will allow for improved statistics and the inclusion of more and faster CMEs from the more active solar cycle 23.

\justify

Even though our investigation focused on 10 years of data from the twin STEREO spacecraft, some of our statistical results are constrained by the limited sample sizes, especially when focusing on sub-categories. For example, there are only 12 sheaths without shocks driven by MEs that are closest to true MCs. However, this dataset with 142 sheaths was enough to shed some light on the importance of the upstream solar wind conditions and the average properties of the driving ME on the average sheath properties. In addition, measurements of the same CME in radial conjunction close to the Sun by Parker Solar Probe \citep{Fox:2016} or Solar Orbiter \citep{Mueller:2013} and at or near 1 AU by \textit{Wind} or STEREO-A will allow us to better understand the evolution of the sheath, especially given that previous work focusing on the radial evolution of MEs and sheaths \citep[e.g.,][]{Winslow:2015,Good:2019,Salman:2020} lacked plasma measurements in the innermost heliosphere.

\begin{acknowledgements}

\justify

T.~M.~S and N.~L acknowledge support from NSF grant AGS1435785 and NASA grants 80NSSC17K0556, 80NSSC20K0431, NNX15AB87G, and 80NSSC20K0700. We also acknowledge the use of Heliospheric shock database, which can be found at \url{http://ipshocks.fi/} and STEREO shock and CME lists, compiled by Dr. Lan~K.~Jian and available at \url{https://stereo-ssc.nascom.nasa.gov/}. We are grateful to the STEREO mission team and NASA/GSFC's Space Physics Data Facility's CDAWeb service (available at \url{https://cdaweb.gsfc.nasa.gov/index.html/}) for providing the data needed for this study. We also thank the anonymous reviewer
for critically reading the manuscript and suggesting substantial improvements.

\end{acknowledgements}

\bibliography{Salman}{}

\begin{thebibliography}{}
\expandafter\ifx\csname natexlab\endcsname\relax\def\natexlab#1{#1}\fi
\providecommand{\url}[1]{\href{#1}{#1}}
\providecommand{\dodoi}[1]{doi:~\href{http://doi.org/#1}{\nolinkurl{#1}}}
\providecommand{\doeprint}[1]{\href{http://ascl.net/#1}{\nolinkurl{http://ascl.net/#1}}}
\providecommand{\doarXiv}[1]{\href{https://arxiv.org/abs/#1}{\nolinkurl{https://arxiv.org/abs/#1}}}

\bibitem[{{Alves} {et~al.}(2016){Alves}, {Da Silva}, {Souza}, {Sibeck},
  {Jauer}, {Vieira}, {Walsh}, {Silveira}, {Marchezi}, {Rockenbach}, {Lago},
  {Mendes}, {Tsurutani}, {Koga}, {Kanekal}, {Baker}, {Wygant}, \&
  {Kletzing}}]{Alves:2016}
{Alves}, L.~R., {Da Silva}, L.~A., {Souza}, V.~M., {et~al.} 2016, Geophys. Res.
  Lett., 43, 978, \dodoi{10.1002/2015GL067066}

\bibitem[{{Burlaga} {et~al.}(1981){Burlaga}, {Sittler}, {Mariani}, \&
  {Schwenn}}]{Burlaga:1981}
{Burlaga}, L., {Sittler}, E., {Mariani}, F., \& {Schwenn}, R. 1981, J. Geophys.
  Res., 86, 6673

\bibitem[{{Chi} {et~al.}(2016){Chi}, {Shen}, {Wang}, {Xu}, {Ye}, \&
  {Wang}}]{Chi:2016}
{Chi}, Y., {Shen}, C., {Wang}, Y., {et~al.} 2016, Solar Phys.,
  \dodoi{10.1007/s11207-016-0971-5}

\bibitem[{{Chree}(1913)}]{Chree:1913}
{Chree}, C. 1913, Philosophical Transactions of the Royal Society of London.
  Series A, Containing Papers of a Mathematical or Physical Character, 212, 75,
  \dodoi{10.1098/rsta.1913.0003}

\bibitem[{{DeForest} {et~al.}(2013){DeForest}, {Howard}, \&
  {McComas}}]{DeForest:2013}
{DeForest}, C.~E., {Howard}, T.~A., \& {McComas}, D.~J. 2013, The Astrophysical
  Journal, 769, 43, \dodoi{10.1088/0004-637X/769/1/43}

\bibitem[{{D{\'e}moulin} \& {Dasso}(2009)}]{Demoulin:2009}
{D{\'e}moulin}, P., \& {Dasso}, S. 2009, Astron. Astrophys., 498, 551

\bibitem[{{D{\'e}moulin} {et~al.}(2008){D{\'e}moulin}, {Nakwacki}, {Dasso}, \&
  {Mandrini}}]{Demoulin:2008}
{D{\'e}moulin}, P., {Nakwacki}, M.~S., {Dasso}, S., \& {Mandrini}, C.~H. 2008,
  Solar Phys., 250, 347, \dodoi{10.1007/s11207-008-9221-9}

\bibitem[{{Echer} {et~al.}(2013){Echer}, {Tsurutani}, \&
  {Gonzalez}}]{Echer:2013}
{Echer}, E., {Tsurutani}, B.~T., \& {Gonzalez}, W.~D. 2013, J. Geophys. Res.,
  118, 385, \dodoi{10.1029/2012JA018086}

\bibitem[{{Farrugia} {et~al.}(1997){Farrugia}, {Burlaga}, \&
  {Lepping}}]{Farrugia:1997}
{Farrugia}, C.~J., {Burlaga}, L.~F., \& {Lepping}, R.~P. 1997, Geophysical
  Monogr. Ser., 98, 91

\bibitem[{{Farrugia} {et~al.}(1991){Farrugia}, {Dunlop}, \&
  {Elliott}}]{Farrugia:1991}
{Farrugia}, C.~J., {Dunlop}, M.~W., \& {Elliott}, S. 1991, Journal of
  Atmospheric and Terrestrial Physics, 53, 1039,
  \dodoi{10.1016/0021-9169(91)90050-H}

\bibitem[{{Fox} {et~al.}(2016){Fox}, {Velli}, \& {Bale}}]{Fox:2016}
{Fox}, N.~J., {Velli}, M.~C., \& {Bale}, S.~D. 2016, Space Science Reviews,
  204, 7, \dodoi{10.1007/s11214-015-0211-6}

\bibitem[{{Galvin} {et~al.}(2008){Galvin}, {Kistler}, {Popecki}, {Farrugia},
  {Simunac}, {Ellis}, {M{\"o}bius}, {Lee}, {Boehm}, {Carroll}, {Crawshaw},
  {Conti}, {Demaine}, {Ellis}, {Gaidos}, {Googins}, {Granoff}, {Gustafson},
  {Heirtzler}, {King}, {Knauss}, {Levasseur}, {Longworth}, {Singer}, {Turco},
  {Vachon}, {Vosbury}, {Widholm}, {Blush}, {Karrer}, {Bochsler}, {Daoudi},
  {Etter}, {Fischer}, {Jost}, {Opitz}, {Sigrist}, {Wurz}, {Klecker}, {Ertl},
  {Seidenschwang}, {Wimmer-Schweingruber}, {Koeten}, {Thompson}, \&
  {Steinfeld}}]{Galvin:2008}
{Galvin}, A.~B., {Kistler}, L.~M., {Popecki}, M.~A., {et~al.} 2008, Space
  Science Reviews, 136, 437, \dodoi{10.1007/s11214-007-9296-x}

\bibitem[{{Gonzalez} {et~al.}(1994){Gonzalez}, {Joselyn}, {Kamide}, {Kroehl},
  {Rostoker}, {Tsurutani}, \& {Vasyliunas}}]{Gonzalez:1994}
{Gonzalez}, W.~D., {Joselyn}, J.~A., {Kamide}, Y., {et~al.} 1994, J. Geophys.
  Res., 99, 5771

\bibitem[{{Good} {et~al.}(2020){Good}, {Ala-Lahti}, {Palmerio}, {Kilpua}, \&
  {Osmane}}]{Good:2020}
{Good}, S.~W., {Ala-Lahti}, M., {Palmerio}, E., {Kilpua}, E.~K.~J., \&
  {Osmane}, A. 2020, The Astrophysical Journal, 893,
  \dodoi{10.3847/1538-4357/ab7fa2}

\bibitem[{{Good} {et~al.}(2019){Good}, {Kilpua}, {LaMoury}, {Forsyth},
  {Eastwood}, \& {M{\"o}stl}}]{Good:2019}
{Good}, S.~W., {Kilpua}, E.~K.~J., {LaMoury}, A.~T., {et~al.} 2019, Journal of
  Geophysical Research: Space Physics, 124, \dodoi{10.1029/2019JA026475}

\bibitem[{{Gopalswamy}(2003)}]{Gopalswamy:2003}
{Gopalswamy}, N. 2003, Advances in Space Research, 31, 869,
  \dodoi{10.1016/S0273-1177(02)00888-8}

\bibitem[{{Gopalswamy}(2006)}]{Gopalswamy:2006}
---. 2006, Space Science Reviews, 124, 145, \dodoi{10.1007/s11214-006-9102-1}

\bibitem[{{Gopalswamy}(2008)}]{Gopalswamy:2008}
---. 2008, J. Atmos. Sol. Terr. Phys., 70, 2078,
  \dodoi{10.1016/j.jastp.2008.06.010}

\bibitem[{{Gopalswamy} {et~al.}(2009){Gopalswamy}, {M{\"a}kel{\"a}}, {Xie},
  {Akiyama}, \& {Yashiro}}]{Gopalswamy:2009}
{Gopalswamy}, N., {M{\"a}kel{\"a}}, P., {Xie}, H., {Akiyama}, S., \& {Yashiro},
  S. 2009, J. Geophys. Res., 114, A00A22, \dodoi{10.1029/2008JA013686}

\bibitem[{{Guo} {et~al.}(2010){Guo}, {Feng}, {Zhang}, {Zuo}, \&
  {Xiang}}]{Guo:2010}
{Guo}, J., {Feng}, X., {Zhang}, J., {Zuo}, P., \& {Xiang}, C. 2010, Journal of
  Geophysical Research: Space Physics, 115, \dodoi{10.1029/2009JA015140}

\bibitem[{{Hietala} {et~al.}(2014){Hietala}, {Kilpua}, {Turner}, \&
  {Angelopoulos}}]{Hietala:2014}
{Hietala}, H., {Kilpua}, E.~K.~J., {Turner}, D.~L., \& {Angelopoulos}, V. 2014,
  Geophys. Res. Lett., 41, 2258, \dodoi{10.1002/2014GL059551}

\bibitem[{{Hudson} {et~al.}(2006){Hudson}, {Bougeret}, \&
  {Burkepile}}]{Hudson:2006}
{Hudson}, H.~S., {Bougeret}, J., \& {Burkepile}, J. 2006, Space Science
  Reviews, 123, 13, \dodoi{10.1007/s11214-006-9009-x}

\bibitem[{{Hudson} {et~al.}(2014){Hudson}, {Baker}, {Goldstein}, {Kress},
  {Paral}, {Toffoletto}, \& {Wiltberger}}]{Hudson:2014}
{Hudson}, M.~K., {Baker}, D.~N., {Goldstein}, J., {et~al.} 2014, Geophys. Res.
  Lett., 41, 1113, \dodoi{10.1002/2014GL059222}

\bibitem[{{Huttunen} \& {Koskinen}(2004)}]{Huttunen:2004}
{Huttunen}, K., \& {Koskinen}, H. 2004, Annales Geophysicae, 22, 1729,
  \dodoi{10.5194/angeo-22-1729-2004}

\bibitem[{{Janvier} {et~al.}(2019){Janvier}, {Winslow}, {Good}, {Bonhomme},
  {D{\'e}moulin}, \& {Dasso}}]{Janvier:2019}
{Janvier}, M., {Winslow}, R.~M., {Good}, S., {et~al.} 2019, Journal of
  Geophysical Research: Space Physics, 124, \dodoi{10.1029/2018JA025949}

\bibitem[{{Jian} {et~al.}(2011){Jian}, {Russell}, \& {Luhmann}}]{Jian:2011}
{Jian}, L.~K., {Russell}, C.~T., \& {Luhmann}, J.~G. 2011, Solar Phys., 274,
  321, \dodoi{10.1007/s11207-011-9737-2}

\bibitem[{{Jian} {et~al.}(2018){Jian}, {Russell}, {Luhmann}, \&
  {Galvin}}]{Jian:2018}
{Jian}, L.~K., {Russell}, C.~T., {Luhmann}, J.~G., \& {Galvin}, A.~B. 2018, The
  Astrophysical Journal, 885, 114, \dodoi{10.3847/1538-4357/aab189}

\bibitem[{{Jones} {et~al.}(2002){Jones}, {Rees}, {Balogh}, \&
  {Forsyth}}]{Jones:2002}
{Jones}, G.~H., {Rees}, A., {Balogh}, A., \& {Forsyth}, R.~J. 2002, Geophys.
  Res. Lett., 11, 1520, \dodoi{10.1029/2001GL014110}

\bibitem[{{Joselyn} \& {Tsurutani}(1990)}]{Joselyn:1990}
{Joselyn}, J.~A., \& {Tsurutani}, B.~T. 1990, EOS Trans., 71, 1808,
  \dodoi{10.1029/90EO00350}

\bibitem[{{Kaiser}(2005)}]{Kaiser:2005}
{Kaiser}, M.~L. 2005, Advances in Space Research, 36, 1483

\bibitem[{{Kamide} {et~al.}(1998){Kamide}, {Yokoyama}, {Gonzalez}, {Tsurutani},
  {Daglis}, {Brekke}, \& {Masuda}}]{Kamide:1998b}
{Kamide}, Y., {Yokoyama}, N., {Gonzalez}, W., {et~al.} 1998, J. Geophys. Res.,
  103, 6917, \dodoi{10.1029/97JA03337}

\bibitem[{{Kataoka} {et~al.}(2005){Kataoka}, {Watari}, {Shimada}, {Shimazu}, \&
  {Marubashi}}]{Kataoka:2005}
{Kataoka}, R., {Watari}, S., {Shimada}, N., {Shimazu}, H., \& {Marubashi}, K.
  2005, Geophys. Res. Lett., 32, L12103, \dodoi{10.1029/2005GL022777}

\bibitem[{{Kaymaz} \& {Siscoe}(2006)}]{Kaymaz:2006}
{Kaymaz}, Z., \& {Siscoe}, G. 2006, Solar Phys., 239, 437,
  \dodoi{10.1007/s11207-006-0308-x}

\bibitem[{{Kilpua} {et~al.}(2017{\natexlab{a}}){Kilpua}, {Balogh}, {von
  Steiger}, \& {Liu}}]{Kilpua:2017a}
{Kilpua}, E.~K.~J., {Balogh}, A., {von Steiger}, R., \& {Liu}, Y.~D.
  2017{\natexlab{a}}, Space Science Reviews, 212, 1271,
  \dodoi{10.1007/s11214-017-0411-3}

\bibitem[{{Kilpua} {et~al.}(2019){Kilpua}, {Fontaine}, {Moissard},
  {Ala-Lahti,}, {Palmerio}, \& {Yordanova}}]{Kilpua:2019b}
{Kilpua}, E.~K.~J., {Fontaine}, D., {Moissard}, C., {et~al.} 2019, Space
  Weather, 17, 1257, \dodoi{10.1029/2019SW002217}

\bibitem[{{Kilpua} {et~al.}(2013){Kilpua}, {Hietala}, {Koskinen}, {Fontaine},
  \& {Turc}}]{Kilpua:2013}
{Kilpua}, E.~K.~J., {Hietala}, H., {Koskinen}, H.~E.~J., {Fontaine}, D., \&
  {Turc}, L. 2013, Annales Geophysicae, 31, 1559,
  \dodoi{10.5194/angeo-31-1559-2013}

\bibitem[{{Kilpua} {et~al.}(2017{\natexlab{b}}){Kilpua}, {Koskinen}, \&
  {Pulkkinen}}]{Kilpua:2017b}
{Kilpua}, E.~K.~J., {Koskinen}, H.~E.~J., \& {Pulkkinen}, T.~I.
  2017{\natexlab{b}}, Living Reviews in Solar Physics, 14,
  \dodoi{10.1007/s41116-017-0009-6}

\bibitem[{{Kilpua} {et~al.}(2015{\natexlab{a}}){Kilpua}, {Lumme}, {Andreeova},
  {Isavnin}, \& {Koskinen}}]{Kilpua:2015a}
{Kilpua}, E.~K.~J., {Lumme}, E., {Andreeova}, K., {Isavnin}, A., \& {Koskinen},
  H.~E.~J. 2015{\natexlab{a}}, Journal of Geophysical Research: Space Physics,
  120, 4112, \dodoi{10.1002/2015JA021138}

\bibitem[{{Kilpua} {et~al.}(2015{\natexlab{b}}){Kilpua}, {Hietala}, {Turner},
  {Koskinen}, {Pulkkinen}, {Rodriguez}, {Reeves}, {Claudepierre}, \&
  {Spence}}]{Kilpua:2015b}
{Kilpua}, E.~K.~J., {Hietala}, H., {Turner}, D.~L., {et~al.}
  2015{\natexlab{b}}, Geophys. Res. Lett., 42, 3076,
  \dodoi{10.1002/2015GL063542}

\bibitem[{{Klein} \& {Burlaga}(1982)}]{Klein:1982}
{Klein}, L.~W., \& {Burlaga}, L.~F. 1982, J. Geophys. Res., 87, 613,
  \dodoi{10.1029/JA087iA02p00613}

\bibitem[{{Lindsay} {et~al.}(1994){Lindsay}, {Russell}, {Luhmann}, \&
  {Gazis}}]{Lindsay:1994}
{Lindsay}, G.~M., {Russell}, C.~T., {Luhmann}, J.~G., \& {Gazis}, P. 1994, J.
  Geophys. Res., 99, \dodoi{10.1029/93JA02666}

\bibitem[{{Lugaz} {et~al.}(2015){Lugaz}, {Farrugia}, {Huang}, \&
  {Spence}}]{Lugaz:2015b}
{Lugaz}, N., {Farrugia}, C.~J., {Huang}, C.~L., \& {Spence}, H.~E. 2015,
  Geophys. Res. Lett., 42, 4694, \dodoi{10.1002/2015GL064530}

\bibitem[{{Lugaz} {et~al.}(2016{\natexlab{a}}){Lugaz}, {Farrugia}, {Huang},
  {Winslow}, {Spence}, \& {Schwadron}}]{Lugaz:2016a}
{Lugaz}, N., {Farrugia}, C.~J., {Huang}, C.-L., {et~al.} 2016{\natexlab{a}},
  Nature Communications, 7, 13001, \dodoi{10.1038/ncomms13001}

\bibitem[{{Lugaz} {et~al.}(2016{\natexlab{b}}){Lugaz}, {Farrugia}, {Winslow},
  {Al-Haddad}, {Kilpua}, \& {Riley}}]{Lugaz:2016}
{Lugaz}, N., {Farrugia}, C.~J., {Winslow}, R.~M., {et~al.} 2016{\natexlab{b}},
  Journal of Geophysical Research: Space Physics, 121,
  \dodoi{10.1002/2016JA023100}

\bibitem[{{Lugaz} {et~al.}(2017){Lugaz}, {Farrugia}, {Winslow}, {Small},
  {Manion}, \& {Savani}}]{Lugaz:2017a}
---. 2017, The Astrophysical Journal, 848, 75, \dodoi{10.3847/1538-4357/aa8ef9}

\bibitem[{{Lugaz} \& {Kintner}(2013)}]{Lugaz:2013a}
{Lugaz}, N., \& {Kintner}, P. 2013, Solar Phys., 285, 47,
  \dodoi{10.1007/s11207-012-9948-1}

\bibitem[{{Lugaz} {et~al.}(2019){Lugaz}, {Winslow}, \& {Farrugia}}]{Lugaz:2019}
{Lugaz}, N., {Winslow}, R.~M., \& {Farrugia}, C.~J. 2019, Journal of
  Geophysical Research: Space Physics, 125, \dodoi{10.1029/2019JA027213}

\bibitem[{{Luhmann} {et~al.}(2008){Luhmann}, {Curtis}, {Schroeder}, {McCauley},
  {Lin}, {Larson}, {Bale}, {Sauvaud}, {Aoustin}, {Mewaldt}, {Cummings},
  {Stone}, {Davis}, {Cook}, {Kecman}, {Wiedenbeck}, {von Rosenvinge}, {Acuna},
  {Reichenthal}, {Shuman}, {Wortman}, {Reames}, {Mueller-Mellin}, {Kunow},
  {Mason}, {Walpole}, {Korth}, {Sanderson}, {Russell}, \&
  {Gosling}}]{Luhmann:2008}
{Luhmann}, J.~G., {Curtis}, D.~W., {Schroeder}, P., {et~al.} 2008, Space
  Science Reviews, 136, 117, \dodoi{10.1007/s11214-007-9170-x}

\bibitem[{{Manchester} {et~al.}(2017){Manchester}, {Kilpua}, {Liu}, {Lugaz},
  {Riley}, {T{\"o}r{\"o}k}, \& {Vr{\v s}nak}}]{Manchester:2017}
{Manchester}, W., {Kilpua}, E.~K.~J., {Liu}, Y.~D., {et~al.} 2017, Space
  Science Reviews, 212, 1159, \dodoi{10.1007/s11214-017-0394-0}

\bibitem[{{Manchester} {et~al.}(2005){Manchester}, {Gombosi}, {De Zeeuw},
  {Sokolov}, {Roussev}, {Powell}, {K{\' o}ta}, {T{\' o}th}, \&
  {Zurbuchen}}]{Manchester:2005}
{Manchester}, W.~B., {Gombosi}, T.~I., {De Zeeuw}, D.~L., {et~al.} 2005, The
  Astrophysical Journal, 622, 1225

\bibitem[{{Mas{\'i}as-Meza} {et~al.}(2016){Mas{\'i}as-Meza}, {Dasso},
  {D{\'e}moulin}, {Rodriguez}, \& {Janvier}}]{Masias:2016}
{Mas{\'i}as-Meza}, J.~J., {Dasso}, S., {D{\'e}moulin}, P., {Rodriguez}, L., \&
  {Janvier}, M. 2016, Astron. Astrophys., 592,
  \dodoi{10.1051/0004-6361/201628571}

\bibitem[{{McComas} {et~al.}(2013){McComas}, {Angold}, {Elliott}, {Livadiotis},
  {Schwadron}, {Skoug}, \& {Smith}}]{McComas:2013}
{McComas}, D.~J., {Angold}, N., {Elliott}, H.~A., {et~al.} 2013, The
  Astrophysical Journal, 779, 10, \dodoi{10.1088/0004-637X/779/1/2}

\bibitem[{{McComas} {et~al.}(1989){McComas}, {Gosling}, {Bame}, {Smith}, \&
  {Cane}}]{McComas:1989}
{McComas}, D.~J., {Gosling}, J.~T., {Bame}, S.~J., {Smith}, E.~J., \& {Cane},
  H.~V. 1989, J. Geophys. Res., 94, 1465, \dodoi{10.1029/JA094iA02p01465}

\bibitem[{{Mitsakou} {et~al.}(2009){Mitsakou}, {Babasidis}, \&
  {Moussas}}]{Mitsakou:2009}
{Mitsakou}, E., {Babasidis}, G., \& {Moussas}, X. 2009, Advances in Space
  Research, 43, 495, \dodoi{10.1016/j.asr.2008.08.003}

\bibitem[{{Mitsakou} \& {Moussas}(2014)}]{Mitsakou:2014}
{Mitsakou}, E., \& {Moussas}, X. 2014, Solar Phys., 289, 3137,
  \dodoi{10.1007/s11207-014-0505-y}

\bibitem[{{M{\"u}ller} {et~al.}(2013){M{\"u}ller}, {Marsden}, {St.~Cyr}, \&
  {Gilbert}}]{Mueller:2013}
{M{\"u}ller}, D., {Marsden}, R.~G., {St.~Cyr}, O.~C., \& {Gilbert}, H.~R. 2013,
  Solar Phys., 285, 25, \dodoi{10.1007/s11207-012-0085-7}

\bibitem[{{Nieves-Chinchilla} {et~al.}(2018){Nieves-Chinchilla}, {Vourlidas},
  \& {Raymond}}]{Nieves:2018}
{Nieves-Chinchilla}, T., {Vourlidas}, A., \& {Raymond}, J.~C. 2018, Solar
  Phys., 293, \dodoi{10.1007/s11207-018-1247-z}

\bibitem[{{Ontiveros} \& {Gonzalez-Esparza}(2010)}]{Ontiveros:2010}
{Ontiveros}, V., \& {Gonzalez-Esparza}, J.~A. 2010, J. Geophys. Res., 115,
  A10244, \dodoi{10.1029/2010JA015471}

\bibitem[{{Owens} {et~al.}(2005){Owens}, {Cargill}, {Pagel}, {Siscoe}, \&
  {Crooker }}]{Owens:2005}
{Owens}, M.~J., {Cargill}, P.~J., {Pagel}, C., {Siscoe}, G.~L., \& {Crooker },
  N.~U. 2005, J. Geophys. Res., 110, A01105, \dodoi{10.1029/2004JA010814}

\bibitem[{{Owens} {et~al.}(2017){Owens}, {Lockwood}, \&
  {Barnard}}]{Owens:2017a}
{Owens}, M.~J., {Lockwood}, M., \& {Barnard}, L.~A. 2017, Scientific Reports,
  7, 4152, \dodoi{10.1038/s41598-017-04546-3}

\bibitem[{{Palmerio} {et~al.}(2016){Palmerio}, {Kilpua}, \&
  {Savani}}]{Palmerio:2016}
{Palmerio}, E., {Kilpua}, E.~K.~J., \& {Savani}, N.~P. 2016, Annales
  Geophysicae, 34, 313, \dodoi{10.5194/angeo-34-313-2016}

\bibitem[{{Phan} {et~al.}(2013){Phan}, {Paschmann}, {Gosling}, {Oieroset},
  {Fujimoto}, {Drake}, \& {Angelopoulos}}]{Phan:2013}
{Phan}, T.~D., {Paschmann}, G., {Gosling}, J.~T., {et~al.} 2013, Geophys. Res.
  Lett., 40, 11, \dodoi{10.1029/2012GL054528}

\bibitem[{{Pulkkinen} {et~al.}(2007){Pulkkinen}, {Partamies}, {Huttunen},
  {Reeves}, \& {Koskinen}}]{Pulkkinen:2007}
{Pulkkinen}, T.~I., {Partamies}, N., {Huttunen}, K.~E.~J., {Reeves}, G.~D., \&
  {Koskinen}, H.~E.~J. 2007, Geophys. Res. Lett., 34, L02105,
  \dodoi{10.1029/2006GL027775}

\bibitem[{{Reames}(1999)}]{Reames:1999}
{Reames}, D.~V. 1999, Space Science Reviews, 90, 413

\bibitem[{{Richardson} \& {Cane}(2010)}]{Richardson:2010}
{Richardson}, I.~G., \& {Cane}, H.~V. 2010, Solar Phys., 264, 189,
  \dodoi{10.1007/s11207-010-9568-6}

\bibitem[{{Richardson} \& {Cane}(2012)}]{Richardson:2012}
---. 2012, J. Space Weather Space Clim, 2, \dodoi{10.1051/swsc/2012001}

\bibitem[{{Richardson} {et~al.}(2001){Richardson}, {Cliver}, \&
  {Cane}}]{Richardson:2001}
{Richardson}, I.~G., {Cliver}, E.~W., \& {Cane}, H.~V. 2001, Geophys. Res.
  Lett., 28, 2569, \dodoi{10.1029/2001GL013052}

\bibitem[{{Rodriguez} {et~al.}(2016){Rodriguez}, {Mas{\'i}as-Meza}, \&
  {Dasso}}]{Rodriguez:2016}
{Rodriguez}, L., {Mas{\'i}as-Meza}, J.~J., \& {Dasso}, S. 2016, Solar Phys.,
  291, 2145, \dodoi{10.1007/s11207-016-0955-5}

\bibitem[{{Russell} \& {Mulligan}(2002)}]{Russell:2002}
{Russell}, C.~T., \& {Mulligan}, T. 2002, Planetary and Space Science, 50, 527,
  \dodoi{10.1016/s0032-0633(02)00031-4}

\bibitem[{{Salman} {et~al.}(2020){Salman}, {Winslow}, \& {Lugaz}}]{Salman:2020}
{Salman}, T.~M., {Winslow}, R.~M., \& {Lugaz}, N. 2020, Journal of Geophysical
  Research: Space Physics, 125, \dodoi{10.1029/2019JA027084}

\bibitem[{{Savani} {et~al.}(2011){Savani}, {Owens}, {Rouillard}, {Forsyth},
  {Kusano}, {Shiota}, {Kataoka}, {Jian}, \& {Bothmer}}]{Savani:2011b}
{Savani}, N.~P., {Owens}, M.~J., {Rouillard}, A.~P., {et~al.} 2011, The
  Astrophysical Journal, 732, 117, \dodoi{10.1088/0004-637X/732/2/117}

\bibitem[{{Shen} {et~al.}(2007){Shen}, {Wang}, {Ye}, {Zhao}, {Gui}, \&
  {Wang}}]{Shen:2007}
{Shen}, C., {Wang}, Y., {Ye}, P., {et~al.} 2007, The Astrophysical Journal,
  670, 849

\bibitem[{{Siscoe} \& {Odstrcil}(2008)}]{Siscoe:2008}
{Siscoe}, G., \& {Odstrcil}, D. 2008, J. Geophys. Res., 113, A00B07,
  \dodoi{10.1029/2008JA013142}

\bibitem[{{Takahashi} \& {Shibata}(2017)}]{Takahashi:2017}
{Takahashi}, T., \& {Shibata}, K. 2017, The Astrophys. Journ. Lett., 837,
  \dodoi{10.3847/2041-8213/aa624c}

\bibitem[{{Tsurutani} {et~al.}(1988){Tsurutani}, {Gonzalez}, {Tang}, \&
  {Akasofu}}]{Tsurutani:1988}
{Tsurutani}, B.~T., {Gonzalez}, W.~D., {Tang}, F., \& {Akasofu}, S.~I.~{Smith},
  E.~J. 1988, J. Geophys. Res., 93, 8519

\bibitem[{{Turner} {et~al.}(2012){Turner}, {Shprits}, {Hartinger}, \&
  {Angelopoulos}}]{Turner:2012}
{Turner}, D.~L., {Shprits}, Y., {Hartinger}, M., \& {Angelopoulos}, V. 2012,
  Nature Physics, 8, 208, \dodoi{10.1038/nphys2185}

\bibitem[{{Vr{\v s}nak} {et~al.}(2013){Vr{\v s}nak}, {{\v Z}ic}, {Vrbanec},
  {Temmer}, {Rollett}, {M{\"o}stl}, {Veronig}, {{\v C}alogovi{\'c}},
  {Dumbovi{\'c}}, {Luli{\'c}}, {Moon}, \& {Shanmugaraju}}]{Vrsnak:2013}
{Vr{\v s}nak}, B., {{\v Z}ic}, T., {Vrbanec}, D., {et~al.} 2013, Solar Phys.,
  285, 295, \dodoi{10.1007/s11207-012-0035-4}

\bibitem[{{Webb} \& {Howard}(2012)}]{Webb:2012}
{Webb}, D.~F., \& {Howard}, T.~A. 2012, Living Reviews in Solar Physics, 9, 3,
  \dodoi{10.12942/lrsp-2012-3}

\bibitem[{{Winslow} {et~al.}(2015){Winslow}, {Lugaz}, {Philpott}, {Schwadron},
  {Farrugia}, {Anderson}, \& {Smith}}]{Winslow:2015}
{Winslow}, R.~M., {Lugaz}, N., {Philpott}, L.~C., {et~al.} 2015, Journal of
  Geophysical Research: Space Physics, 120, 6101, \dodoi{10.1002/2015JA021200}

\bibitem[{{Wu} \& {Lepping}(2016)}]{Wu:2016}
{Wu}, C.~C., \& {Lepping}, R.~P. 2016, Solar Phys., 291, 265,
  \dodoi{10.1007/s11207-015-0806-9}

\bibitem[{{Xiang} {et~al.}(2017){Xiang}, {Tu}, {Li}, {Ni}, {Morley}, \&
  {Baker}}]{Xiang:2017}
{Xiang}, Z., {Tu}, W., {Li}, X., {et~al.} 2017, Journal of Geophysical
  Research: Space Physics, 122, 9858, \dodoi{10.1002/2017JA024487}

\bibitem[{{Yermolaev} {et~al.}(2010){Yermolaev}, {Nikolaeva}, {Lodkina}, \&
  {Yermolaev}}]{Yermolaev:2010}
{Yermolaev}, Y.~I., {Nikolaeva}, N.~S., {Lodkina}, I.~G., \& {Yermolaev}, M.~Y.
  2010, Annales Geophysicae, 28, 2177, \dodoi{10.5194/angeo-28-2177-2010}

\bibitem[{{Zhang} {et~al.}(2007){Zhang}, {Richardson}, {Webb}, {Gopalswamy},
  {Huttunen}, {Kasper}, {Nitta}, {Poomvises}, {Thompson}, {Wu}, {Yashiro}, \&
  {Zhukov}}]{Zhang:2007}
{Zhang}, J., {Richardson}, I.~G., {Webb}, D.~F., {et~al.} 2007, J. Geophys.
  Res., 112, A10102, \dodoi{10.1029/2007JA012321}

\end{thebibliography}
\bibliographystyle{aasjournal}

\end{document}